\documentclass[a4paper,11pt]{article}
\pdfoutput=1
\usepackage{jheppub} 
\usepackage{graphicx}
\usepackage{amsmath}
\usepackage{amssymb}
\usepackage{bbold}
\usepackage{float}
\usepackage{slashed}
\usepackage{dcolumn}
\usepackage{bm}
\usepackage{color}
\usepackage{multirow}
\usepackage{hhline}
\usepackage{changepage}
\usepackage{caption}

\usepackage{array} 
\newcolumntype{C}[1]{>{\centering\arraybackslash}p{#1}}

\usepackage{booktabs} 

\newcommand{\lsim}{{\;\raise0.3ex\hbox{$<$\kern-0.75em\raise-1.1ex\hbox{$\sim$}}\;}}
\newcommand{\gsim}{{\;\raise0.3ex\hbox{$>$\kern-0.75em\raise-1.1ex\hbox{$\sim$}}\;}}
\def\bea{\begin{eqnarray}}
\def\eea{\end{eqnarray}}
\def\bec{\begin{center}}
\def\ec{\end{center}}

\def\beq{\begin{equation}}
\def\eeq{\end{equation}}

\def\bea{\begin{eqnarray}}
\def\eea{\end{eqnarray}}
\def\beq#1\eeq{\begin{align}#1\end{align}}
\def\beqnn#1\eeq{\begin{align*}#1\end{align*}}
\def\ba{\begin{array}}
\def\ea{\end{array}}
\def\bc{\begin{center}}
\def\ec{\end{center}}

\newcommand{\dis}[1]{\begin{equation}\begin{split}#1\end{split}\end{equation}}

\preprint{CTPU-PTC-26-16}

\title{The EDM inverse problem: Identifying the sources of  CP violation and PQ breaking with  electric dipole moments}
 
 \author{
    Kiwoon Choi\footnote{email: kchoi@ibs.re.kr} and
    Sang Hui Im\footnote{email: imsanghui@ibs.re.kr}
 }
     
\affiliation{
Particle Theory and Cosmology Group, Center for Theoretical Physics of the Universe, \\
 Institute for Basic Science (IBS), Daejeon 34126, Korea \
    }

\abstract{
Many extensions of the Standard Model (SM) generically introduce new sources of CP violation, which can induce observable $P$-odd and $T$-odd permanent electric dipole moments (EDMs) of nuclei, atoms, and molecules. A future observation of nonvanishing EDMs would therefore provide a sensitive probe of physics beyond the SM, while also posing a nontrivial inverse problem: identifying their underlying ultraviolet origin.
In this work, we identify six representative classes of CP-violating effective operators near the QCD scale, including the QCD $\theta$-term, that are particularly relevant for low-energy EDMs and can arise in a broad range of SM extensions. We show that these operator classes lead to distinct EDM patterns across different systems, thereby enabling discrimination among them through experimentally measured EDMs.
We further emphasize that EDM measurements can shed light on the origin of the vacuum expectation value of the QCD axion. In particular, they may help distinguish whether a nonzero axion vacuum expectation value is predominantly induced by high-scale Peccei--Quinn symmetry-breaking effects, such as those associated with quantum gravity, or by the interplay between beyond-the-SM CP violation and the QCD anomaly.
}

\vspace{4cm}

\begin{document} 
\maketitle
\flushbottom
 
\section{Introduction}

Despite its tremendous success, the Standard Model (SM) of particle physics requires extension, as it does not contain a viable dark matter candidate and cannot account for baryogenesis (see, e.g., \cite{Trodden:1998ym} for a review) or neutrino masses. Physics beyond the Standard Model (BSM) that addresses these problems generically introduces new sources of CP violation. Such CP violation can be transmitted to the SM sector, giving rise to sizable $P$-odd and $T$-odd permanent electric dipole moments (EDMs) of nuclei, atoms, and molecules \cite{Chupp:2017rkp, Cai:2017jrq, Okawa:2019arp}. Roughly speaking, BSM physics at energy scales ranging from the TeV to the PeV scale can induce EDMs that may be observable in near-future experiments \cite{Pospelov:2025vzj}.

Once nonvanishing EDMs are observed, a central question will be the identification of the underlying ultraviolet (UV) physics responsible for the observed signals. In principle, properties of the underlying UV physics may be inferred, at least partially, from the available EDM data. We refer to the problem of identifying the specific underlying physics responsible for a given set of EDM measurements as \emph{the EDM inverse problem}.
This problem has been investigated in a number of earlier works, both in terms of CP-violating (CPV) operators at the hadronic scale \cite{Chupp:2014gka, Chupp:2017rkp, Gaul:2023hdd, Degenkolb:2024eve} and at higher energy scales \cite{deVries:2011an, Dekens:2014jka, deVries:2018mgf, deVries:2021sxz, Choi:2023bou}.

In this work, we revisit the EDM inverse problem. We first identify six important classes of CPV operators around the QCD scale ($\sim 1~\mathrm{GeV}$) that are particularly relevant for low-energy EDMs and can arise from broad classes of well-motivated BSM scenarios. These include the QCD $\theta$-term, which already exists in the SM; three additional classes of hadronic operators associated with the EDMs and chromo-EDMs of light quarks and gluons; and two classes of (semi-)leptonic operators, including the electron EDM.
We then demonstrate that the ratios of EDMs of nuclei, atoms, and molecules predicted by these operator classes exhibit distinctive patterns, depending on which class of CPV operators dominates. Our approach is similar in spirit to previous studies \cite{deVries:2011an, Dekens:2014jka, deVries:2018mgf, deVries:2021sxz, Choi:2023bou}, but we present updated and more quantitative results by incorporating state-of-the-art theoretical inputs for atomic and molecular EDMs, as well as hadronic observables induced by CPV operators at the QCD scale.

An intriguing aspect of EDM physics is its connection to the Peccei--Quinn (PQ) mechanism, which addresses the strong CP problem via the QCD axion \cite{Peccei:1977hh, Weinberg:1977ma, Wilczek:1977pj}. As discussed in Refs.~\cite{deVries:2018mgf, deVries:2021sxz, Choi:2023bou}, patterns of nuclear and atomic EDMs can provide valuable information about the underlying mechanism responsible for resolving the strong CP problem.
Hadronic BSM CPV operators generically induce large quantum corrections to the QCD $\theta$ parameter \cite{Morozov:1984goy, Chang:1991hz, deVries:2018mgf, deVries:2021sxz}, thereby spoiling ultraviolet solutions to the strong CP problem, such as the Nelson--Barr mechanism \cite{Nelson:1983zb, Barr:1984qx}. Therefore, if experimentally observed EDM ratios are consistent with predictions from hadronic BSM CPV operators while being inconsistent with those from the QCD $\theta$-dominated scenario, this would point toward the necessity of an infrared solution to the strong CP problem, such as the PQ mechanism, to explain the smallness of $\theta$ \cite{deVries:2018mgf}.

In our previous work \cite{Choi:2023bou}, we generalized and reinterpreted this observation by allowing for the possibility of a simply fine-tuned $\theta$, and showed that certain classes of hadronic CPV operators, with or without the PQ mechanism, can be distinguished by their characteristic EDM ratios. In particular, we argued that EDM measurements may reveal whether a nonzero vacuum expectation value (VEV) of the QCD axion predominantly originates from PQ-symmetry breaking at high scales, such as quantum-gravitational effects, or from the interplay between hadronic BSM CPV operators and the QCD anomaly. In other words, EDM data may provide insight into the dominant PQ-breaking source beyond the QCD anomaly, which is crucial for understanding the axion quality problem (see, e.g., \cite{Dine:2022mjw} for a recent review).
In the present work, we extend this analysis by including (i) all potentially important BSM CPV operators arising from well-motivated scenarios, (ii) EDMs of light nuclei and diamagnetic atoms targeted by ongoing and planned experiments, and (iii) scenarios in which multiple classes of CPV operators contribute comparably to the observed EDMs.

We find that EDM measurements of light nuclei, such as the neutron, proton, deuteron, and helion, are particularly powerful for discriminating among the four classes of hadronic operators, including the $\theta$-term and the EDMs and chromo-EDMs of quarks and gluons. This observation strongly motivates storage-ring experiments that can directly measure the EDMs of charged light nuclei \cite{pEDM:2025nlu}. In contrast, EDM data from diamagnetic atoms with heavy nuclei, such as $^{199}$Hg, $^{129}$Xe, and $^{171}$Yb, may provide suggestive information about the UV source but are unlikely to be conclusive due to large theoretical uncertainties. Octupole-deformed systems, such as $^{225}$Ra, may offer improved sensitivity in this respect, although still less than that of light nuclei. We also discuss how EDM measurements of two different paramagnetic molecules, such as HfF$^+$ and ThO, can disentangle the electron EDM from the electron--nucleon CPV coupling as previously analyzed in \cite{Fleig:2018bsf}, and we further explore the implications for hadronic CPV sources.

It should be emphasized that many of the matching relations used to connect CPV operators at the QCD scale to experimentally measured EDMs are subject to intrinsic uncertainties that cannot be reliably quantified with the current level of theoretical understanding of the relevant  QCD, nuclear and atomic physics effects. Our results should therefore be interpreted with these uncertainties in mind. Our primary goal is to assess how far one can proceed using the currently available theoretical inputs, while also motivating future theoretical efforts toward a more quantitative understanding of the matching relations relevant for EDMs.

This paper is organized as follows. In Sec.~\ref{sec:UVsrc}, we identify six classes of CPV operators around the QCD scale that are particularly relevant for low-energy EDMs and arise from broad classes of well-motivated BSM scenarios. We also discuss the radiative corrections to these operators from the BSM scale down to the QCD scale, as well as the impact of certain hadronic operators on the PQ mechanism that solves the strong CP problem via a QCD axion field whose vacuum expectation value can be identified with the QCD $\theta$ parameter. In Sec.~\ref{sec:IRsrc}, we describe the leading infrared CPV operators at the hadronic level below the QCD scale and present updated results for their matching to the CPV sources at the QCD scale. In Sec.~\ref{sec:edms}, we summarize state-of-the-art calculations of EDMs for light nuclei, atoms, and molecules, together with experimental prospects. Section~\ref{sec:iden} contains the main results of this work, where we demonstrate that the six classes of CPV operators at the QCD scale, either with or without the PQ mechanism, can be distinguished using appropriate combinations of EDM data. We conclude in Sec.~\ref{sec:conc}.

\section{UV sources for CP violation and PQ breaking} \label{sec:UVsrc}

Permanent electric dipole moments (EDMs) of elementary particles and nucleons arise from CP-violating (CPV) interactions among Standard Model (SM) particles. In general, gauge-invariant CPV interactions of SM fields above the electroweak scale can be written as
\dis{
{\cal L}_{\rm CPV}(\mu > m_W) 
= 
{\cal L}_{\rm KM} 
+ \frac{g_s^2}{32\pi^2} \bar{\theta} \, G^a_{\mu \nu} \widetilde{G}^{a\mu \nu}  
+ {\cal L}_6 
+ \cdots \,,
}
where ${\cal L}_{\rm KM}$ denotes the charged-current weak interactions containing the Kobayashi--Maskawa (KM) phase, $\bar{\theta}$ is the QCD vacuum angle defined in the basis where the quark mass matrices are real and diagonal, and ${\cal L}_6$ represents a set of CPV dimension-six operators constructed from SM fields. 
The ellipsis includes the dimension-five Weinberg operator responsible for Majorana neutrino masses, as well as operators of dimension greater than six. We neglect these contributions, as they are suppressed either by small neutrino masses or by additional powers of the heavy mass scale.
We also neglect the contribution of ${\cal L}_{\rm KM}$ to EDMs, since it is highly suppressed. This suppression arises because EDMs are flavor-diagonal observables, whereas CP violation in ${\cal L}_{\rm KM}$ is encoded in flavor-changing interactions~(see, e.g., \cite{Pospelov:2025vzj}).

A complete classification of operators in ${\cal L}_6$ is given in Ref.~\cite{Grzadkowski:2010es}. However, at the QCD scale $\mu \sim 1~\mathrm{GeV}$, many of these operators are integrated out, leaving the following operators that provide the leading contributions to EDMs:
\bea
{\cal L}_\theta &=& \frac{g_s^2}{32\pi^2}  \bar{\theta} \, G^a_{\mu \nu} \widetilde{G}^{a \mu \nu}, \label{L1}\\
{\cal L}_{\rm dipole} &=&  \frac{1}{3} w f^{abc} G_{\alpha}^{a \mu}  G_\mu^{b\delta} \widetilde{G}_\delta^{c\alpha}
-\frac{i}{2} \sum_{q=u, d, s} \tilde{d}_q \, g_s \, \bar{q} \sigma^{\mu \nu} G_{\mu \nu} \gamma_5 q
-\frac{i}{2} \sum_{f=u, d, s, e} d_f  \, \bar{f} \sigma^{\mu \nu} F_{\mu \nu} \gamma_5 f , \nonumber \\ \label{L2} \\
{\cal L}_{\rm 4\text{-}fermi} &=& \left(i\sum_{D=d, s, b} \mathrm{Im}(C_{e e DD}) \,(e_L^\dagger e_R)\, (\bar{D}_R D_L)+ \mathrm{h.c.}\right) + \cdots, \label{L3}
\eea
where  ${\cal L}_{\rm dipole}$ includes the Weinberg three-gluon operator, which can be interpreted as a gluonic chromo-electric dipole moment (CEDM), the light-quark CEDMs, and the EDMs of light quarks and the electron. The ellipsis in ${\cal L}_{\rm 4\text{-}fermi}$ denotes CPV four-quark operators and additional (semi-)leptonic operators that are not considered in this work.

The SM and various beyond-the-SM (BSM) scenarios can be characterized by a small number of dominant operators among those appearing in Eqs.~(\ref{L1})--(\ref{L3}). In the SM, the QCD ${\theta}$-term ${\cal L}_\theta$ provides the dominant source of EDMs, if nonzero EDMs are observed in the near future, since contributions from the KM phase are far below current experimental sensitivities. 
In contrast, one or more dipole operators in ${\cal L}_{\rm dipole}$ can dominate in well-motivated BSM scenarios, such as supersymmetric extensions of the SM~\cite{Ibrahim:1997gj, Chang:1998uc, Abel:2001vy}, two-Higgs-doublet models (2HDMs)~\cite{Jung:2013hka}, and universal theories~\cite{Barbieri:2004qk, Cirigliano:2019vfc}, in which the BSM sector communicates with the SM primarily through gauge interactions or the Higgs portal.

The semi-leptonic four-fermion operators in ${\cal L}_{\rm 4\text{-}fermi}$ can be particularly important in the minimal supersymmetric Standard Model (MSSM) or the type-II 2HDM in the large $\tan\beta$ regime, where their effects are enhanced by the large value of  $\tan\beta$~\cite{hep-ph/0204359}. 
By contrast, other types of CPV four-fermion operators, including four-quark operators, are typically suppressed by the small SM Yukawa couplings of light fermions in such BSM scenarios~\cite{Jung:2013hka, Pospelov:2005pr, Ellis:2008zy}. Certain four-quark operators can play an important role in left--right symmetric models~\cite{Dekens:2014jka, deVries:2021sxz}, which we do not consider here.

We therefore focus on six potentially important classes of UV CPV parameters arising from the SM and various BSM scenarios:
\dis{
X_i^{\rm UV} \in \left\{ \bar{\theta}, w, \tilde{d}_q, d_q, d_e, \mathrm{Im}(C_{e e DD}) \right\}, \label{Xuv}
}
where $q=u,d,s$ and $D=d,s,b$. Parameters within the same class typically give comparable contributions to EDMs, while those from different classes can differ by orders of magnitude.

In many BSM scenarios, chirality violation in quark (C)EDMs originates from quark masses, implying a relation between strange- and down-quark (C)EDMs,
\dis{
\tilde{d}_s = \frac{m_s}{m_d} \tilde{d}_d, \quad d_s = \frac{m_s}{m_d} d_d. \label{strange}
}
By contrast, the up-quark (C)EDM is not generally related to the down-quark (C)EDM by a simple mass ratio, due to differences in electric charges and the $\tan\beta$ dependence of Yukawa couplings for up- and down-type quarks. Accordingly, we will often impose Eq.~(\ref{strange}) while treating $d_u$ ($\tilde{d}_u$) and $d_d$ ($\tilde{d}_d$) as independent parameters.

The different classes of CPV parameters in Eq.~(\ref{Xuv}) are related to each other in some cases. Below, we discuss two important effects that lead to nontrivial relations among them.

\subsection{QCD axion effect}

If the Peccei--Quinn (PQ) mechanism is realized to solve the strong CP problem,
the QCD $\theta$ term is replaced by the axion coupling to gluons,
\bea
\frac{g_s^2}{32\pi^2}
\frac{a}{f_a} G^a_{\mu\nu}\tilde G^{a\mu\nu},
\eea
where $a(x)$ denotes the axion field with decay constant $f_a$.
This coupling arises from the breaking of the PQ symmetry $U(1)_{\rm PQ}$
by the QCD anomaly and generates the standard QCD axion potential,
\bea
V_{\rm QCD}(a) \sim m_\pi^2 f_\pi^2
\cos\!\left(\frac{a}{f_a}\right).
\eea
It also implies that the QCD $\theta$ parameter is determined by the vacuum
expectation value (VEV) of the axion field as
\bea
\bar{\theta} = \frac{\langle a \rangle}{f_a}.
\eea

If the axion potential is entirely dominated by the QCD contribution
$V_{\rm QCD}$, minimizing the potential yields $\bar{\theta}=0$, thereby solving
the strong CP problem.
On the other hand, in the presence of CP-odd effective interactions at the QCD
scale,
\bea
{\cal L}_{\rm BSM} = \sum_i \lambda_i {\cal O}_i,
\label{bsm_effective}
\eea
which may arise from BSM physics at higher scales, the axion potential
receives an additional contribution estimated as~\cite{Choi:2023bou}
\dis{
\delta V_{\rm BSM}(a) \sim
\sum_i \lambda_i \int d^4x
\left\langle
\frac{g_s^2}{32\pi^2} G \widetilde{G}(x)\, {\cal O}_i(0)
\right\rangle_{a=0}
\frac{a}{f_a}
+ {\cal O}\!\left(\frac{a}{f_a}\right)^2 .
}
This additional contribution shifts the axion VEV away from the
CP-conserving point, leading to a nonzero effective QCD $\theta$ parameter,
\dis{
\bar{\theta}_{\rm BSM}
= \frac{\langle a \rangle_{\rm BSM}}{f_a}
\sim
\frac{
\sum_i \lambda_i \int d^4x
\left\langle
\frac{g_s^2}{32\pi^2} G \widetilde{G}(x)\, {\cal O}_i(0)
\right\rangle_{a=0}
}{
f_\pi^2 m_\pi^2
}.
}
To solve the strong CP problem, this induced $\theta$ parameter must satisfy
the experimental bound $|\bar{\theta}_{\rm BSM}| \lesssim 10^{-10}$, which imposes
stringent constraints on the Wilson coefficients $\lambda_i$ of the
BSM-induced effective interactions in Eq.~(\ref{bsm_effective}).

Among the operators listed in Eqs.~(\ref{L1})--(\ref{L3}), quark chromo-electric dipole moments (CEDMs) and the gluon CEDM (the Weinberg three-gluon operator) can generate sizable contributions to $\bar{\theta}_{\rm BSM}$. The axion VEV induced by quark CEDMs has been computed using QCD sum rules in Ref.~\cite{Pospelov:1999rg}, whereas no dedicated calculation exists for the gluon CEDM contribution. Employing na\"ive dimensional analysis (NDA) for the latter, one finds
\dis{
\bar{\theta}_{\rm BSM}
= \frac{\langle a \rangle_{\rm BSM}}{f_a}
=
\frac{m_0^2}{2}
\sum_{q=u,d,s} \frac{\tilde{d}_q}{m_q}
+ {\cal O}(4\pi f_\pi^2 w), \label{thetapq}
}
where
$m_0^2 \equiv g_s \langle \bar{q} G^{\mu\nu} \sigma_{\mu\nu} q\rangle / \langle \bar{q} q\rangle
= 0.8(1)\,\mathrm{GeV}^2$~\cite{Belyaev:1982sa}.

Another important source of a nonzero axion VEV arises from PQ-symmetry
breaking effects other than the QCD anomaly, such as those induced by quantum
gravity. Since global symmetries are generally expected to be violated by
quantum gravity, the PQ symmetry associated with the QCD axion is expected to
be broken near the quantum gravity scale~\cite{Kim:1988ix, Rey:1989mg, Barr:1992qq, Kamionkowski:1992mf, Holman:1992us, Ghigna:1992iv, Kallosh:1995hi}. For example, string or brane instantons in
string theory, as well as gravitational wormholes, can generate an axion
potential of the form (see, e.g., \cite{Blumenhagen:2009qh, Hebecker:2018ofv}
for reviews)
\dis{
\delta V_{\rm UV}
=
\Lambda_{\rm UV}^4 e^{-S_{\rm ins}}
\cos\!\left(\frac{a}{f_a}+\delta_{\rm UV}\right),
}
where $\Lambda_{\rm UV}$ is a model-dependent UV scale, $S_{\rm ins}$ is the
Euclidean instanton action, and $\delta_{\rm UV}$ is a CP-violating phase
generically of order unity. This potential induces an axion VEV
\dis{
\bar{\theta}_{\rm UV}
=
\frac{\langle a \rangle_{\rm UV}}{f_a}
\sim
e^{-S_{\rm ins}}
\frac{\Lambda_{\rm UV}^4 \sin\delta_{\rm UV}}{f_\pi^2 m_\pi^2},
}
which can naturally satisfy $|\bar{\theta}_{\rm UV}| \lesssim 10^{-10}$ for
moderately large values of $S_{\rm ins}$.

In general, the total axion VEV receives contributions from both sources,
\dis{
\bar{\theta}_{\rm PQ}
\equiv \frac{\langle a\rangle}{f_a}
=
\bar{\theta}_{\rm UV}
+
\bar{\theta}_{\rm BSM}. \label{thetapq1}
}
Therefore, in the presence of a QCD axion, the effective $\bar{\theta}$ parameter is nontrivially related to the quark CEDMs $\tilde{d}_q$ and the gluon CEDM $w$. As discussed later, this relation may be experimentally testable through EDM measurements if the UV contribution $\bar{\theta}_{\rm UV}$ is subdominant compared to $\bar{\theta}_{\rm BSM}$. Conversely, EDM data may provide insight into the dominant origin of a nonvanishing axion VEV.

\subsection{Renormalization effect} \label{RGE}

If the operators in Eqs.~(\ref{L1})--(\ref{L3}) are generated at a high scale well above the QCD scale, they undergo significant mixing through renormalization-group (RG) evolution down to $\mu\sim 1~\mathrm{GeV}$. The relevant RG equations are given by~\cite{Morozov:1984goy, Braaten:1990gq, Chang:1991hz, Degrassi:2005zd, Hisano:2012cc}
\bea
\frac{d{\bf K}}{d\ln\mu}
=
\frac{g_s^2}{16\pi^2}\,\gamma\,{\bf K}, \label{RGECs}
\eea
where ${\bf K}\equiv (K_1,\,K_2,\,K_3)^T$ with
\bea
K_1(\mu) = \frac{d_q(\mu)}{m_q Q_q}, \qquad
K_2(\mu) = \frac{\tilde{d}_q(\mu)}{m_q}, \qquad
K_3(\mu) = \frac{w(\mu)}{g_s}. \label{K123}
\eea
The anomalous-dimension matrix $\gamma$ is
\bea
\gamma =
\begin{pmatrix}
\gamma_e & \gamma_{eq} & 0 \\
0 & \gamma_q & \gamma_{Gq} \\
0 & 0 & \gamma_G
\end{pmatrix}
=
\begin{pmatrix}
8C_F & 8C_F & 0 \\
0 & 16C_F - 4N_c & -2N_c \\
0 & 0 & N_c + 2n_f + \beta_0
\end{pmatrix},
\eea
where $C_F=(N_c^2-1)/(2N_c)=4/3$, $N_c=3$ is the number of colors, $n_f$ is the number of active light quark flavors with $m_q<\mu$, and $\beta_0=(33-2n_f)/3$.

The strong coupling constant and the quark masses evolve according to
\dis{
\frac{d\alpha_s}{d\ln\mu}
=
-2\beta_0 \frac{\alpha_s^2}{4\pi},
\qquad
\frac{dm_q}{d\ln\mu}
=
-8\frac{\alpha_s}{4\pi} m_q. \label{alphasmq}
}
For renormalization scales $\mu<m_c$ and a BSM scale $\Lambda\geq 1~\mathrm{TeV}$ at which the CPV operators are generated, analytic solutions to Eq.~(\ref{RGECs}) in terms of $g_s(m_q)$ ($q=t,b,c$) are obtained in Ref.~\cite{Choi:2023bou}. Using numerical values for $g_s(m_q)$, one finds
\dis{\hskip -2cm
w(1~\mathrm{GeV})
\simeq
0.33
\left(\frac{g_s(\Lambda)}{g_s(1~\mathrm{TeV})}\right)^{15/7}
w(\Lambda), \label{wrun}
}
\dis{
\frac{\Delta\tilde{d}_q}{m_q}(1~\mathrm{GeV})
\simeq
\left[
0.19\left(\frac{g_s(\Lambda)}{g_s(1~\mathrm{TeV})}\right)^{1/3}
-
0.06\left(\frac{g_s(\Lambda)}{g_s(1~\mathrm{TeV})}\right)^{15/7}
\right] w(\Lambda), \label{dqtrun}
}
where $\Delta\tilde{d}_q$ denotes the RG-induced contribution to the quark CEDM arising from the gluon CEDM.

Consequently, depending on the scale $\Lambda$ at which the gluon CEDM is generated,
\dis{
\frac{\Delta\tilde{d}_q}{m_q}(1~\mathrm{GeV})
=
r(\Lambda)\,
w(1~\mathrm{GeV}), \label{rgdqtw}
}
with representative values
\dis{
r(\Lambda)\simeq
\begin{cases}
0.41, & \Lambda = 1~\mathrm{TeV},\\
0.53, & \Lambda = 10~\mathrm{TeV},\\
0.65, & \Lambda = 100~\mathrm{TeV}.
\end{cases} \label{Lvalues}
}
This RG mixing plays an important role in the gluon CEDM contribution to CPV pion--nucleon interactions, which are key ingredients in nuclear and atomic EDMs.

\section{CP-odd interactions at IR} \label{sec:IRsrc}

The CP-odd operators of the previous section generated at a BSM scale induce CP-odd interactions of hadrons and electrons at low energies below the QCD scale.  
At leading order, the relevant operators are\footnote{In principle, the three-pion operator $\Delta_\pi \pi^0 \pi^+ \pi^-$ can also be included. In our previous analysis~\cite{Choi:2023bou}, however, we found its contribution to nuclear and atomic EDMs to be generally negligible, and we therefore ignore it here. We also neglect the nuclear spin-dependent interactions $\bar{e} \sigma^{\mu \nu} e \,\bar{N} S_\mu v_\nu N$ and $\bar{e}e \,\partial^\mu \bar{N} S_\mu N$, since they are typically subleading compared with the nuclear spin-independent interaction $(\bar{e} i \gamma_5 e)(\bar{N} N)$.}
\bea
{\cal L}_{\rm dipole} &=&  -\frac{i}{2}\, \bar{N} \left(d_p \frac{1+\tau_3}{2} + d_n \frac{1-\tau_3}{2} \right) \sigma^{\mu \nu} F_{\mu \nu} \gamma_5 N
-\frac{i}{2}\, d_e \,\bar{e}\,\sigma^{\mu \nu} F_{\mu \nu} \gamma_5 e,  \label{neEDMs}\\
{\cal L}_{\pi N} &=&   \bar{g}_0 \,\bar{N}\,\vec{\tau}\!\cdot\!\vec{\pi}\, N + \bar{g}_1 \,\pi_3 \,\bar{N} N, \label{piN}\\
{\cal L}_{4N} &=&   C_1 \,\bar{N}N\, D_\mu(N^\dagger S^{\mu} N)
+ C_2 \,\bar{N} \vec{\tau} N \cdot D_\mu (N^\dagger \vec{\tau} S^\mu N), \label{NR4N} \\
{\cal L}_{e N} &=&  -\frac{G_F}{\sqrt{2}} \,(\bar{e} i \gamma_5 e)\,
\bar{N} (C_S^{(0)} + C_S^{(1)} \tau_3)N, \label{eN}
\eea
where $N = (p~ n)^T$ are the nucleon fields, $\vec{\pi} = (\pi_1, \pi_2, \pi_3)$ are the pion fields, $\tau_i\,(i=1,2,3)$ are Pauli matrices in isospin space, $\sigma_i\,(i=1,2,3)$ are Pauli matrices in spin space, and $S^\mu=(0,\vec{\sigma}/2)$.
The EDMs of nucleons and electrons in Eq. (\ref{neEDMs}) directly contribute to nuclear, atomic, or molecular EDMs, while the CP-odd interactions in Eqs. (\ref{piN})-(\ref{eN}) contribute to the EDMs by polarizing nuclei, atoms, or molecules. We will discuss in details how these CP-odd operators contribute to the EDMs of nuclei, atoms, and molecules in Sec. \ref{sec:edms}. In this section, let us discuss the connection between these IR CPV operators and the UV CPV operators of Sec. \ref{sec:UVsrc}.

In Eqs. (\ref{neEDMs})-(\ref{eN}) we have nine independent IR CPV parameters which are 
\dis{
X_i^{\rm IR} \in \left\{ d_p, d_n, \bar{g}_0, \bar{g}_1, C_1, C_2, d_e, C_S^{(0)}, C_S^{(1)} \right\}. \label{XIR}
}
The parameters $X_i^{\rm IR}$ originate from the UV CPV parameters $X_i^{\rm UV}$ in Eq. (\ref{Xuv}), and they are matched at a scale $\mu^*$ near the QCD scale $\sim 1$ GeV by linear relations approximately: 
\dis{
X_i^{\rm IR} = \sum_j M_{ij} X_j^{\rm UV} \qquad \text{at} \qquad \mu=\mu^* .
}
In the following we discuss the matrix elements $M_{ij}$ and the choice of the matching scale $\mu^*$.

\subsection{Nucleon electric dipole moments}

The nucleon EDMs $d_N\, (N=p, n)$ can be sizably induced from the non-leptonic sources $X_i^{\rm UV}\in\{\bar{\theta},\tilde d_q,d_q, w\}$ $(q=u,d,s)$ among the parameters in Eq. (\ref{Xuv}).
Let us first discuss contributions from
$\{\bar{\theta},\tilde d_q,d_q\}$, and then turn to the contribution from $w$. Here we will largely rely on the calculations using QCD sum rules for nucleon EDMs \cite{Pospelov:1999ha, Pospelov:2000bw, Demir:2002gg, Hisano:2012sc, Hisano:2015rna, Haisch:2019bml, Yamanaka:2020kjo}.

The nucleon EDMs induced by $\{\bar{\theta},\tilde d_q,d_q\}$ were computed by QCD sum rules in
Refs.~\cite{Pospelov:1999ha, Pospelov:2000bw, Hisano:2012sc, Hisano:2015rna}:
\dis{
d_N(\bar{\theta}, \tilde{d}_q, d_q) = c_N \,\Theta_N(\bar{\theta}, \tilde{d}_q, d_q), \qquad (N=p,n), \label{dN_CEDM}
}
where $c_N$ is an overall normalization factor that depends on the single-pole contribution to the two-point correlator of the nucleon interpolating field. The functions $\Theta_N$ are given by
\dis{
\Theta_p(\bar{\theta}, \tilde{d}_q, d_q) =\,&
\chi m_* \left[
(4 e_u - e_d)\left(\bar{\theta}-\frac{m_0^2}{2} \frac{\tilde{d}_s}{m_s}\right)
+\frac{m_0^2}{2}(\tilde{d}_u-\tilde{d}_d)\left(\frac{4e_u}{m_d}+\frac{e_d}{m_u}\right)
\right]
\\
&+\frac{1}{8}(2\kappa+\xi)(4e_u \tilde{d}_u -e_d \tilde{d}_d) + (4d_u-d_d),
\\
\Theta_n(\bar{\theta}, \tilde{d}_q, d_q) =\,&
\chi m_* \left[
(4 e_d - e_u)\left(\bar{\theta}-\frac{m_0^2}{2} \frac{\tilde{d}_s}{m_s}\right)
+\frac{m_0^2}{2}(\tilde{d}_d-\tilde{d}_u)\left(\frac{4e_d}{m_u}+\frac{e_u}{m_d}\right)
\right]
\\
&+\frac{1}{8}(2\kappa+\xi)(4e_d \tilde{d}_d -e_u \tilde{d}_u) + (4d_d-d_u),
\label{thepn}
}
where $m_* \equiv \left(\sum_{q=u,d,s} m_q^{-1}\right)^{-1}\simeq m_um_d/(m_u+m_d)$ and $e_q$ denotes the electromagnetic charge of the quark $q$.
The expressions involve quark-condensate susceptibilities~\cite{Pospelov:2000bw},
\dis{
\langle \bar{q} \sigma_{\mu \nu} q \rangle = e_q \chi F_{\mu \nu} \langle \bar{q} q \rangle,\quad
g_s\langle \bar{q} G_{\mu \nu} q \rangle = e_q \kappa F_{\mu \nu} \langle \bar{q} q \rangle, \\
g_s \langle \bar{q} G^{\mu \nu} \sigma_{\mu \nu} q \rangle = m_0^2 \langle \bar{q} q \rangle,\quad
2g_s\langle \bar{q} \gamma_5 \widetilde{G}_{\mu \nu} q \rangle = i e_q \xi F_{\mu \nu} \langle \bar{q} q \rangle,
}
whose numerical values are taken as\footnote{Ref.~\cite{Kaneta:2023wrl} adopts smaller values of $\chi$ based on a recent lattice determination, by a factor of $\sim 3$ compared to earlier estimates. As also noted there, there are arguments suggesting that the product $m_0^2\chi$ is approximately fixed, $m_0^2\chi\simeq 6$~\cite{Cata:2009fd}, which supports the original estimate. Further lattice studies are needed; here we use the original value of $\chi$.}
\dis{
&\chi=-5.7(6)\,\mathrm{GeV}^{-2},\qquad
m_0^2=0.8(1)\,\mathrm{GeV}^2,\\
&\kappa=-0.34(10),\qquad
\xi=-0.74(20).
}

The overall factor $c_N$ in Eq.~(\ref{dN_CEDM}) carries a large theoretical uncertainty due to the poorly known single-pole contribution, leading to a range of choice for the Borel mass in the sum-rule analysis. The corresponding uncertainty in nucleon EDMs can exceed $50\%$~\cite{Hisano:2012sc}. Ref.~\cite{Kaneta:2023wrl} instead fixes the normalization by matching to lattice results for quark EDM contributions, which have smaller uncertainties; we follow the same strategy here.
The quark-EDM contributions to nucleon EDMs have been determined with good accuracy in lattice QCD~\cite{Park:2025rxi, Alexandrou:2024ozj}:
\dis{
d_p(d_q) &= g_T^u d_d + g_T^d d_u + g_T^s d_s,\\
d_n(d_q) &= g_T^u d_u + g_T^d d_d + g_T^s d_s,
\label{qEDM_lat}
}
where the tensor charges are\footnote{Ref.~\cite{Vecchi:2025jbb} argues, based on several considerations, that $|g_T^s|$ should be as large as $\mathcal{O}(0.1)\,|g_T^{u,d}|$ barring accidental cancellations, making questions on both the lattice results for $g_T^s$ and the validity of leading-order QCD sum-rule estimates. In this work we nevertheless adopt the lattice values and leading-order sum-rule expressions. A significantly larger $g_T^s$ could modify our estimates of nuclear and atomic EDMs sourced by quark EDMs.}
\dis{
g_T^u = -0.195(16),\qquad
g_T^d = 0.782(28),\qquad
g_T^s = -0.0016(12),
}
given at $\mu=2~\mathrm{GeV}$ in Ref.~\cite{Park:2025rxi}.
To use these to determine $c_N$ in Eq.~(\ref{dN_CEDM}), we evolve the tensor charges to $\mu=1~\mathrm{GeV}$, the matching scale used in Eq.~(\ref{dN_CEDM}) above. We find
$g_T^q(1~\mathrm{GeV})/g_T^q(2~\mathrm{GeV}) \simeq 1.047$, and thus obtain\footnote{One may alternatively take $c_N^{\rm lattice}\simeq \frac14 g_T^d(1~\mathrm{GeV})=0.205(7)$, which has a smaller uncertainty. To be conservative, we choose $g_T^u$ instead.}
\dis{
c_N^{\rm lattice} \simeq - g_T^u(1~\mathrm{GeV}) = 0.204(17).
}
Using this value in Eq.~(\ref{dN_CEDM}), we obtain\footnote{We use the $\overline{\mathrm{MS}}$ quark masses in Ref.~\cite{ParticleDataGroup:2024cfk}: $m_u=2.16(7)$~MeV, $m_d=4.70(7)$~MeV, and $m_s=93.5(8)$~MeV at $\mu=2$~GeV. We evolve them to $\mu=1$~GeV using one-loop QCD running, which gives $m_q(1~\mathrm{GeV})\simeq 1.15\,m_q(2~\mathrm{GeV})$.}
\bea
d_p &=& -1.16(16) \cdot 10^{-16} \bar{\theta} \,e \,\textrm{cm} + e \left(-0.38(6) \tilde{d}_u +0.27(5) \tilde{d}_d +  0.022(4) \tilde{d}_s   \right)  \label{dplat}\\
&& +0.82(7) d_u -0.204(17) d_d\,, \nonumber \\
d_n &=& 0.77(10) \cdot 10^{-16} \bar{\theta} \,e \,\textrm{cm} + e \left(-0.30(6) \tilde{d}_u +0.37(6) \tilde{d}_d -  0.0145(27) \tilde{d}_s   \right) \label{dnpat}\\
&& -0.204(17) d_u +0.82(7) d_d\,. \nonumber 
\eea
Assuming $\tilde{d}_s=(m_s/m_d)\tilde{d}_d$ as discussed around Eq. (\ref{strange}), the above equations become
\bea
d_p &=& -1.16(16) \cdot 10^{-16} \bar{\theta} \,e \,\textrm{cm} + e \left(-0.38(6) \tilde{d}_u +0.70(9) \tilde{d}_d   \right)  \label{dplat'}\\
&& +0.82(7) d_u -0.204(17) d_d\,, \nonumber \\
d_n &=& 0.77(10) \cdot 10^{-16} \bar{\theta} \,e \,\textrm{cm} + e \left(-0.30(6) \tilde{d}_u +0.08(8) \tilde{d}_d   \right) \label{dnpat'}\\
&& -0.204(17) d_u +0.82(7) d_d\,. \nonumber 
\eea
Here we note that $\partial d_n/\partial \tilde d_d$  becomes significantly smaller than
$\partial d_p/\partial \tilde d_d$,
 due to an accidental cancellation between the down- and strange-quark CEDM contributions to $d_n$ under the relation $\tilde{d}_s=(m_s/m_d)\tilde{d}_d$.

If the PQ mechanism for the strong CP problem is realized, leading to Eq.~(\ref{thetapq1}), one finds
\dis{
\Theta_p^{\rm PQ}(\bar{\theta}_{\rm UV}, \tilde{d}_q, d_q) =\,
\chi m_* (4 e_u - e_d)\,\bar{\theta}_{\rm UV}
+\left(\frac{1}{8}(2\kappa+\xi)+\frac12\chi m_0^2\right)(4e_u\tilde{d}_u-e_d\tilde{d}_d)
+(4d_u-d_d), \\
\Theta_n^{\rm PQ}(\bar{\theta}_{\rm UV}, \tilde{d}_q, d_q) =\,
\chi m_* (4 e_d - e_u)\,\bar{\theta}_{\rm UV}
+\left(\frac{1}{8}(2\kappa+\xi)+\frac12\chi m_0^2\right)(4e_d\tilde{d}_d-e_u\tilde{d}_u)
+(4d_d-d_u),
\label{thepnpq}
}
where the strange-quark contribution cancels exactly against $\bar{\theta}_{\rm BSM}$ in Eq.~(\ref{thetapq}).
Numerically,
\bea
d_p^{\rm PQ} &=& -1.16(16) \cdot 10^{-16} \bar{\theta}_{\rm UV} \,e \,\textrm{cm} + e \left(-1.34(23) \tilde{d}_u -0.167(29) \tilde{d}_d   \right) \label{dplatPQ}\\
&& +0.82(7) d_u -0.204(17) d_d\,, \nonumber \\
d_n^{\rm PQ}  &=& 0.77(10) \cdot 10^{-16} \bar{\theta}_{\rm UV} \,e \,\textrm{cm} + e \left(0.33(6) \tilde{d}_u +0.67(12) \tilde{d}_d  \right) \label{dnlatPQ}\\
&& -0.204(17) d_u +0.82(7) d_d\,. \nonumber 
\eea

In the leading sum-rule expressions of Eq. (\ref{thepn}) and Eq. (\ref{thepnpq}), one can observe that the dependence on the quark-condensate susceptibilities largely cancels in the EDM ratio $d_p/d_n$. Moreover, the ratio does not depend on the normalization factor $c_N$. Within the QCD sum-rule approach, therefore, the EDM ratio $d_p/d_n$ is predicted quite precisely. If the nucleon EDMs are dominantly from $\bar{\theta}$, the ratio is
\dis{
\frac{d_p(\bar{\theta})}{d_n(\bar{\theta})}
=
\frac{d_p^{\rm PQ}(\bar{\theta}_{\rm UV})}{d_n^{\rm PQ}(\bar{\theta}_{\rm UV})}
=
\frac{4e_u-e_d+\cdots}{4e_d-e_u+\cdots}
= -\frac{3}{2}(1\pm{\cal O}(0.1))\,.
}
where the ellipsis denotes the higher-order contributions in the operator product expansion (OPE) of the sum rule, which is estimated to be ${\cal O}(10)$\% \cite{Hisano:2012sc}.
In the case that the nucleon EDMs are dominantly from quark CEDMs $\tilde{d}_q$,
assuming the relation $\tilde{d}_s=(m_s/m_d)\tilde{d}_d$ in Eq. (\ref{strange}),
\dis{
\frac{d_p(\tilde{d}_q)}{d_n(\tilde{d}_q)}
&=
\frac{
\left(m_* \left(\frac{4e_u}{m_d}+\frac{e_d}{m_u}\right) + e_u \zeta\right)\tilde{d}_u
-
\left(m_*\left(\frac{8e_u-e_d}{m_d}+\frac{e_d}{m_u}\right)+\frac{e_d}{4}\zeta\right)\tilde{d}_d +\cdots
}{
-\left(m_*\left(\frac{4e_d}{m_u}+\frac{e_u}{m_d}\right)+\frac{e_u}{4}\zeta\right)\tilde{d}_u
+
\left(m_*\left(\frac{4e_d}{m_u}+\frac{2e_u-4e_d}{m_d}\right)+e_d\zeta\right)\tilde{d}_d +\cdots
}\,,
\\
&\simeq
\frac{0.81(6)\tilde{d}_u - 1.51(4)\tilde{d}_d}{0.64(4)\tilde{d}_u - 0.18(4)\tilde{d}_d}  \left( 1\pm {\cal O}(0.1)\right),
}
where $\zeta\equiv (2\kappa+\xi)/(m_0^2\chi)\simeq 0.31(8)$, and the ellipsis again denotes the higher-order contributions in the OPE accounted for by the factor $(1\pm {\cal O}(0.1))$ in the last expression. 
We again note that
$\partial d_n/ \partial \tilde d_d$ is suppressed relative to $\partial d_p/ \partial \tilde d_d$ by about an
order of magnitude, since
\bea
\left(\frac{4e_d}{m_u}+\frac{2e_u-4e_d}{m_d}\right)
\propto (m_d - 2m_u) \approx 0.
\eea
This suppression originates from an accidental cancellation between the
down- and strange-quark contributions to $d_n$ under the relation
$\tilde{d}_s = (m_s/m_d)\tilde{d}_d$.
In the presence of the PQ mechanism, the proton-to-neutron EDM ratio
induced by quark CEDMs can be significantly modified:
\bea
\frac{d_p^{\rm PQ}(\tilde{d}_q)}{d_n^{\rm PQ}(\tilde{d}_q)}
&=&
-\frac{4e_u\tilde{d}_u-e_d\tilde{d}_d+ \cdots}{e_u\tilde{d}_u-4e_d\tilde{d}_d+ \cdots}
=
-\frac{8\tilde{d}_u+\tilde{d}_d}{2\tilde{d}_u+4\tilde{d}_d}  \left( 1\pm {\cal O}(0.1)\right).
\eea
Finally, if the dominant UV CPV source is quark EDMs $d_q$, the ratio turns out to be
\dis{
\frac{d_p(d_q)}{d_n(d_q)}=\frac{d_p^{\rm PQ}(d_q)}{d_n^{\rm PQ}(d_q)}
= -\frac{4d_u-d_d}{d_u-4d_d}  \left( 1\pm {\cal O}(0.1)\right)
}
with the higher-order contributions of the OPE estimated as  ${\cal O}(10)$\% \cite{Hisano:2012sc}.

Let us now discuss the contribution of the gluon CEDM (or Weinberg three-gluon operator) to nucleon EDMs. 
Since the gluon CEDM does not break the chiral symmetry, it turns out that the one-particle reducible contribution from the CP-odd nucleon mass is comparable to the one-particle irreducible contribution unlike the QCD $\theta$-term or quark CEDMs \cite{Bigi:1991rh, Bigi:1990kz}. Therefore, the total contribution has been computed as a sum of the two parts:
\dis{
d_N(w) = d_N^{\rm (red)}(w) + d_N^{\rm (irr)}(w).
}
The reducible contribution is obtained by the chiral rotation of the anomalous magnetic moment with the CP-odd nucleon mass that can be evaluated using QCD sum rules \cite{Demir:2002gg, Haisch:2019bml}:
\dis{
d_N^{\rm (red)}(w)
= -\mu_N^{\rm an}\,\frac{3g_s m_0^2}{32\pi^2}\,w\,
\ln\!\left(\frac{M^2}{\mu_{\rm IR}^2}\right),
}
where $\mu_N^{\rm an}$ is the nucleon anomalous magnetic moment, $g_s$ and $w$ are evaluated at $\mu=1~\mathrm{GeV}$, $M$ is the Borel mass, and $\mu_{\rm IR}$ is an IR cutoff in the sum-rule analysis. The dominant uncertainty arises from $m_0^2$ and the range of $M/\mu_{\rm IR}$, taken as $\sqrt{2}\le M/\mu_{\rm IR}\le 2\sqrt{2}$~\cite{Haisch:2019bml}, yielding
\bea
d_p^{\rm (red)}(w) &=& -23(11)\, w\, e\,\mathrm{MeV},\\
d_n^{\rm (red)}(w) &=&25(11)\, w\, e\,\mathrm{MeV}.
\eea
On the other hand, the irreducible contribution is estimated in Ref.~\cite{Yamanaka:2020kjo} using a nonrelativistic quark model:
\bea
d_p^{\rm (irr)}(w) &=& 4.5(5)\, w\, e\,\mathrm{MeV},\\
d_n^{\rm (irr)}(w) &=& -4.5(5)\, w\, e\,\mathrm{MeV},
\eea
where the uncertainty reflects model-dependent choices for the quark interaction. Combining the two contributions gives, at $\mu=1~\mathrm{GeV}$,
\bea
d_p(w) &=& -19(11)\, w\, e\,\mathrm{MeV},\\
d_n(w) &=& 20(11)\, w\, e\,\mathrm{MeV}.
\eea
Although the absolute normalization carries an $\mathcal{O}(50\%)$ uncertainty, the ratio is significantly more robust within this framework:
\dis{
\frac{d_p(w)}{d_n(w)} = -0.90(3).
}

Finally we remark that the PQ mechanism would make a negligible change for the nucleon EDMs from the gluon CEDM, unlike the quark CEDMs, if the shifted axion VEV from the gluon CEDM does not significantly differ from the NDA estimation in Eq. (\ref{thetapq}).

\subsection{CP-odd pion-nucleon interactions}
The CP-odd pion--nucleon couplings $\bar{g}_0$ and $\bar{g}_1$ in
Eq.~(\ref{piN}) can receive sizable contributions from purely hadronic
UV sources $X_i^{\rm UV} \in \left\{ \bar{\theta}, \tilde{d}_q, w \right\}$
with $q=u,d,s$. The state-of-the-art results for these couplings are
summarized in detail in Ref.~\cite{Choi:2023bou}. Here we briefly present
the essential formulas.

The couplings $\bar{g}_0$ and $\bar{g}_1$ induced by
the QCD $\theta$ term can be computed in chiral perturbation theory
in terms of quark-mass contributions to baryon masses
\cite{Bsaisou:2014zwa, deVries:2015una}\footnote{For comparison, see also large-$N_c$ estimation in \cite{Richardson:2025rnm} which shows good agreement with previous results.}:
\bea
\bar{g}_0 (\bar{\theta})  &=&
\frac{\delta m_N}{2f_\pi} \frac{1-\epsilon^2}{2\epsilon} \bar{\theta}
= 15.7(1.7)\times10^{-3} \, \bar{\theta}, \label{gbar0_th}\\
\bar{g}_1 (\bar{\theta}) &=&
\left(8c_1 m_N \frac{\epsilon(1-\epsilon^2)}{16f_\pi m_N}
\frac{m_\pi^4}{m_K^2 - m_\pi^2}
+ {\cal O}\!\left(\epsilon \frac{m_\pi^4}{m_N^3 f_\pi}\right)\right)
\bar{\theta}
= -3.4(2.4)\times10^{-3} \, \bar{\theta},
\label{gbar1_th}
\eea
where $\delta m_N = m_n - m_p = 2.49(17)$ MeV,
$\epsilon = (m_d-m_u)/(2\bar{m}) = 0.37(3)$,
$\bar{m} = (m_u+m_d)/2 = 3.37(8)$ MeV,
$c_1 = 1.0(3)\, \mathrm{GeV}^{-1}$ is related to the nucleon sigma term
as $\sigma_{\pi N} = -4c_1 m_\pi^2 + {\cal O}(m_\pi^3)$
\cite{Baru:2011bw, Bernard:1996gq},
$f_\pi = 92.2$ MeV, $m_K = 495$ MeV, and $m_\pi = 135$ MeV.

The contributions from quark CEDMs were first computed using QCD sum rules
\cite{Pospelov:2001ys} and have recently been refined within chiral
perturbation theory \cite{deVries:2021sxz}, building on the observation of
Ref.~\cite{Seng:2018wwp}.\footnote{Recently, Ref.~\cite{Bhattacharya:2025blb} revisited the observation made in Ref.~\cite{Seng:2018wwp} and pointed out that the nucleon form factors neglected in Ref.~\cite{Seng:2018wwp} may not be small, based on large-$N_c$ arguments. Estimating these form factors using large-$N_c$ scaling with $N_c=3$ suggests that the previous estimates of $g_0(\tilde{d}_q)$ and $g_1(\tilde{d}_q)$ in Refs.~\cite{Seng:2018wwp, deVries:2021sxz} may receive sizable corrections. We find that, under this rough $N_c=3$ estimate, the corrections can exceed $100\%$ for $\bar{g}_0(\tilde{d}_q)$ and amount to about $70\%$ for $\bar{g}_1(\tilde{d}_q)$. However, since these form factors are currently unknown apart from such large-$N_c$ estimates, we do not include these potential contributions in this work. For the analysis in Sec.~\ref{sec:iden}, the omitted contributions could approximately double the error bars of the predicted EDM ratios sourced by $\tilde{d}_q$, mainly due to the increased uncertainty in $\bar{g}_1(\tilde{d}_q)$, while the impact of $\bar{g}_0(\tilde{d}_q)$ remains small.} The resulting expressions are
\bea
\bar{g}_0 (\tilde{d}_q) &\simeq&
\frac{m_0^2}{8f_\pi}
\left[
\delta_{g_0} \frac{d \delta m_N}{d (\bar{m}\epsilon)}(\tilde{d}_u+\tilde{d}_d)
-\frac{(1-\epsilon^2)\delta m_N}{\epsilon}
\sum_{q=u,d,s}\frac{\tilde d_q}{m_q}
\right], \nonumber \\
&\simeq&
-0.4(0.9)\, \tilde{d}_u\, \mathrm{GeV}
+1.0(0.8)\, \tilde{d}_d\, \mathrm{GeV}
-0.059(10)\, \tilde{d}_s\, \mathrm{GeV},
\label{gbar0_dqt}\\
&\simeq&
-0.4(0.9)\, \tilde{d}_u\, \mathrm{GeV}
-0.2(0.8)\, \tilde{d}_d\, \mathrm{GeV}
~ (\text{for } \tilde{d}_s=(m_s/m_d)\tilde{d}_d),
\label{gbar0_dqt'}\\
\bar{g}_0^{\rm PQ} (\tilde{d}_q) &\simeq&
\delta_{g_0} \frac{m_0^2}{8f_\pi}
\frac{d \delta m_N}{d (\bar{m}\epsilon)}(\tilde{d}_u+\tilde{d}_d)
\simeq 2.2(0.7)\, (\tilde{d}_u + \tilde{d}_d)\, \mathrm{GeV},
\label{gbar0PQ_dqt} \\
\bar{g}_1 (\tilde{d}_q)\simeq \bar{g}_1^{\rm PQ}(\tilde{d}_q)
&\simeq&
\delta_{g_1} \frac{1}{2f_\pi}(\tilde{d}_u-\tilde{d}_d)
\frac{m_0^2}{2} \frac{\sigma_{\pi N}}{\bar{m}}
= 38(13)\, (\tilde{d}_u -\tilde{d}_d)\, \mathrm{GeV},
\label{gbar1_dqt}
\eea
evaluated at the matching scale $\mu = 1$ GeV. Here
$d \delta m_N/d (\bar{m}\epsilon) \simeq \delta m_N/(\bar{m}\epsilon)$,
$\sigma_{\pi N} = 59.1(3.5)$ MeV, and
$\delta_{g_0,g_1} = 1.0 \pm 0.3$ account for theoretical uncertainties.
The superscript ``PQ'' in Eqs.~(\ref{gbar0PQ_dqt}) and (\ref{gbar1_dqt})
denotes the couplings in the presence of the PQ mechanism, including the
effect of the shifted axion VEV in Eq.~(\ref{thetapq}).

Lastly, the contribution of the gluon CEDM to the coupling $\bar{g}_1$
has been computed using chiral perturbation theory combined with QCD
sum rules in Ref.~\cite{Osamura:2022rak}:
\dis{
\bar{g}_1 (w) \simeq \bar{g}_1^{\rm PQ}(w)
\simeq \langle 0| {\cal L}_w | \pi^0 \rangle
\left(\frac{\sigma_{\pi N}}{f_\pi^2 m_\pi^2}
+ \frac{5 g_A^2 m_\pi}{64\pi f_\pi^4} \right)
\simeq \pm (2.6\pm1.5) \times 10^{-3} \, w\, \textrm{GeV}^2,
\label{gbar1_w}
}
evaluated at the matching scale $\mu = 1$ GeV. Here
${\cal L}_w = \frac{1}{3} w f^{abc} G_{\alpha}^{a \mu}
G_\mu^{b\delta} \widetilde{G}_\delta^{c\alpha}$
denotes the Weinberg three-gluon operator (i.e.\ the gluon CEDM), and
$g_A = 1.27$ is the nucleon axial-vector coupling. The sign ambiguity
arises from the QCD sum-rule estimate of the matrix element of the
Weinberg operator.

On the other hand, there is no dedicated computation of $\bar{g}_0(w)$
to date. Applying naive dimensional analysis (NDA), we estimate
\bea
\bar{g}_0(w) \sim \bar{g}_0^{\rm PQ}(w)
\sim \left.(m_u + m_d)\right|_{\mu_{\rm NDA}}
\Lambda_\chi \left. w \right|_{\mu_{\rm NDA}}
\sim 9\times10^{-3} \left. w \right|_{1 \, \mathrm{GeV}}
\, \textrm{GeV}^2,
\label{gbar0_w}
\eea
where $\mu_{\rm NDA} \simeq 225$ MeV is the NDA matching scale defined by
$\alpha_s(\mu_{\rm NDA})/(4\pi) \simeq 1/6$, at which the QCD one-loop
beta function becomes comparable to the two-loop contribution
\cite{Weinberg:1989dx}. The final expression is obtained by evolving the
parameters to $\mu = 1$ GeV for comparison with quantities evaluated
using QCD sum rules.

If the gluon CEDM is generated at a high scale, the quark CEDMs induced
through RG mixing (cf.~Eq.~(\ref{dqtrun})) can provide an important
contribution. Using Eq.~(\ref{rgdqtw}), we obtain
\bea
\bar{g}_1 ( \Delta \tilde{d}_q ) \simeq \bar{g}_1^{\rm PQ} ( \Delta \tilde{d}_q )
&\simeq& -0.11(4)\, r(\Lambda)\, w \,\mathrm{GeV}^2,\\
\bar{g}_0 ( \Delta \tilde{d}_q )
&\simeq& -2(6)\times 10^{-3}\, r(\Lambda)\, w \,\mathrm{GeV}^2,\\
\bar{g}_0^{\rm PQ} ( \Delta \tilde{d}_q )
&\simeq& 17(6)\times 10^{-3}\, r(\Lambda)\, w \,\mathrm{GeV}^2,
\eea
where $\Delta \tilde{d}_q$ denotes the RG-induced quark CEDM generated
from a gluon CEDM at the scale $\Lambda$, and $w$ is renormalized at
$\mu = 1$ GeV. For $\Lambda \gtrsim 1$ TeV, one typically has
$r(\Lambda) = \mathcal{O}(1)$ (see Eq.~(\ref{Lvalues})). It then follows that
\dis{
\bar{g}_1^{(\mathrm{PQ})}(\Delta \tilde{d}_q)
\gg
\bar{g}_1^{(\mathrm{PQ})}(w)
\sim
\bar{g}_0^{(\mathrm{PQ})}(w)
\sim
\bar{g}_0^{(\mathrm{PQ})}(\Delta \tilde{d}_q),
}
at $\mu = 1$ GeV for $\Lambda \gtrsim 1$ TeV, where
$\bar{g}_{0,1}^{(\mathrm{PQ})}(w)$ denote the direct contributions of $w$
in Eqs.~(\ref{gbar1_w}) and (\ref{gbar0_w}). This implies that the
CP-odd pion--nucleon couplings $\bar{g}_{0,1}$ induced by a gluon CEDM
generated at $\Lambda \gtrsim 1$ TeV are dominated by the contributions
arising from RG-induced quark CEDMs at $\mu \sim 1$ GeV.

We now consider the ratios $e\bar{g}_0/(\Lambda_\chi d_n)$ and
$e\bar{g}_1/(\Lambda_\chi d_n)$, which are useful for assessing the
relative sizes of nuclear and atomic EDMs with respect to the neutron
EDM. Throughout, we assume $\tilde{d}_s = (m_s/m_d)\tilde{d}_d$, as in
Eq.~(\ref{strange}).

If the QCD $\theta$ term is the dominant CPV source, we find
\bea
\frac{e \bar{g}_0(\bar{\theta})}{\Lambda_\chi d_n(\bar{\theta})}
&=&
\frac{e \bar{g}_0^{\rm PQ}(\bar{\theta}_{\rm UV})}
{\Lambda_\chi d_n^{\rm PQ}(\bar{\theta}_{\rm UV})}
\left(\sim 4\pi\right)_{\rm NDA}
\simeq 3.5(6),\\
\frac{e \bar{g}_1(\bar{\theta})}{\Lambda_\chi d_n(\bar{\theta})}
&=&
\frac{e \bar{g}_1^{\rm PQ}(\bar{\theta}_{\rm UV})}
{\Lambda_\chi d_n^{\rm PQ}(\bar{\theta}_{\rm UV})}
\left(\sim 4\pi \frac{m_u-m_d}{m_s}\right)_{\rm NDA}
\simeq -0.8(5),
\eea
where the NDA estimates are shown in parentheses (see Appendix for details).

For quark CEDMs, we obtain
\bea
\frac{e \bar{g}_0(\tilde{d}_q)}{\Lambda_\chi d_n(\tilde{d}_q)}
\left(\sim 4\pi\right)_{\rm NDA}
&\simeq&
\frac{4(11)\tilde{d}_u + 2(10)\tilde{d}_d}
{4.4(8)\tilde{d}_u - 1.2(11)\tilde{d}_d},\\
\frac{e \bar{g}_0^{\rm PQ}(\tilde{d}_q)}{\Lambda_\chi d_n^{\rm PQ}(\tilde{d}_q)}
\left(\sim 4\pi \frac{\tilde{d}_u+\tilde{d}_d}{\tilde{d}_q}\right)_{\rm NDA}
&\simeq&
\frac{5.6(2.2)(\tilde{d}_u + \tilde{d}_d)}
{\tilde{d}_u + 2\tilde{d}_d},\\
\frac{e \bar{g}_1(\tilde{d}_q)}{\Lambda_\chi d_n(\tilde{d}_q)}
\left(\sim 4\pi \frac{\tilde{d}_u-\tilde{d}_d}{\tilde{d}_q}\right)_{\rm NDA}
&\simeq&
\frac{-483(165)(\tilde{d}_u - \tilde{d}_d)}
{4.4(8)\tilde{d}_u - 1.2(11)\tilde{d}_d},\\
\frac{e \bar{g}_1^{\rm PQ}(\tilde{d}_q)}{\Lambda_\chi d_n^{\rm PQ}(\tilde{d}_q)}
\left(\sim 4\pi \frac{\tilde{d}_u-\tilde{d}_d}{\tilde{d}_q}\right)_{\rm NDA}
&\simeq&
\frac{99(37)(\tilde{d}_u - \tilde{d}_d)}
{\tilde{d}_u + 2\tilde{d}_d}.
\eea

If the dominant CPV source is a gluon CEDM generated at  $\Lambda$,
we obtain
\bea
\frac{e \bar{g}_0(w)}{\Lambda_\chi d_n(w)}
&\simeq&
-0.08(28)\, r(\Lambda)
\pm \frac{m_u+m_d}{f_\pi},\\
\frac{e \bar{g}_0^{\rm PQ}(w)}{\Lambda_\chi d_n^{\rm PQ}(w)}
&\simeq&
0.7(5)\, r(\Lambda)
\pm \frac{m_u+m_d}{f_\pi},
\eea
\dis{
\frac{e \bar{g}_1(w)}{\Lambda_\chi d_n(w)}
\simeq
\frac{e \bar{g}_1^{\rm PQ}(w)}{\Lambda_\chi d_n^{\rm PQ}(w)}
\simeq
-4.8(3.3)\, r(\Lambda)
\pm 0.11(9),
}
where the terms proportional to $r(\Lambda)$ arise from the RG-induced
quark CEDMs.

\subsection{CP-odd four-nucleon contact interactions}

The CP-odd four-nucleon couplings $C_1$ and $C_2$ in
Eq.~(\ref{NR4N}) contribute to nuclear and atomic EDMs by inducing
nuclear polarization through CP-odd nuclear forces. Their relative
importance can be assessed by comparing the short-range contact
CP-odd nuclear force generated by $C_{1,2}$ with the pion-exchange
CP-odd nuclear force induced by $\bar{g}_{0,1}$. From a comparison of
the corresponding tree-level diagrams, one finds that the contact
contribution can dominate over the pion-exchange contribution only if
\dis{
C_{1,2} \gtrsim \frac{\bar{g}_{0,1}}{m_\pi^2 \Lambda_\chi}
\sim \bar{g}_{0,1}\,\mathrm{fm}^3\,.
}
As discussed in Ref.~\cite{Choi:2023bou}, naive dimensional analysis
(NDA) indicates that this condition can be satisfied only if the dominant
CP-violating (CPV) source is the gluon CEDM among the hadronic UV sources
$X_i^{\rm UV} \in \left\{ \bar{\theta}, \tilde{d}_q, w \right\}$.
The underlying reason is that the gluon CEDM preserves light-quark chiral
symmetry, so that $C_{1,2}(w)$ are not suppressed by light-quark masses,
whereas the QCD $\theta$ term and quark CEDMs break chiral symmetry
and therefore lead to such suppressions.
Therefore, within NDA, the contributions of $C_{1,2}$ to nuclear and
atomic EDMs need to be included only if the dominant CPV source involves
the gluon CEDM.

The calculation of $C_1$ induced by the gluon CEDM beyond NDA has been
performed in Ref.~\cite{Yamanaka:2022qlu} using non-relativistic quark
models. The result is
\dis{
C_1(w) = 0.3^{+5}_{-1.3}\, w \, \mathrm{GeV}^{-1},
\label{C1w}
}
evaluated at the matching scale $\mu = 1$ GeV. On the other hand, there
is as yet no calculation of $C_2(w)$ beyond NDA. Nevertheless, $C_2(w)$
may be negligible compared to $C_1(w)$ if their relative sizes follow
those of their CP-even counterparts:
\bea
{\cal L}_{NNNN} =
-\frac12 k_S \bar{N} N \bar{N} N
+ 2k_T \bar{N}\vec{\sigma} N \cdot \bar{N}\vec{\sigma} N.
\eea
It is well known that the CP-even couplings are
$k_S = -120.8\,\mathrm{GeV}^{-2}$ and
$k_T = 1.8\,\mathrm{GeV}^{-2}$ \cite{Epelbaum:2008ga}. The corresponding
CP-odd couplings are given by
\footnote{The non-relativistic Lagrangian (\ref{NR4N}) can be derived from
the relativistic Lagrangian
\dis{
{\cal L}_{4N} =
-\frac{m_N}{2} \left[
C_1 \bar{N} N \bar{N} i\gamma_5 N
+ C_2 \bar{N} \vec{\tau} N \cdot \bar{N} \vec{\tau} i\gamma_5 N
\right],
}
which, using Fierz identities, can be rewritten as
\dis{
{\cal L}_{4N} =
-\frac{m_N}{2} \left[
(C_1 - 2C_2) \bar{N} N \bar{N} i\gamma_5 N
- C_2 \bar{N} \vec{\sigma} N \cdot \bar{N} \vec{\sigma} i\gamma_5 N
\right].
}
}
\bea
\bar{k}_S = m_N (C_1 - 2 C_2), \qquad
\bar{k}_T = \frac{1}{4} m_N C_2,
\eea
which implies $|C_2| \ll |C_1|$ if $|\bar{k}_T| \ll |\bar{k}_S|$.

Finally, we estimate the ratio $e f_\pi^2 C_1(w)/d_n(w)$, which will be
useful in Sec.~\ref{sec:iden}:
\dis{
\frac{e f_\pi^2 C_1(w)}{d_n(w)}
\left(\sim 1 \right)_{\rm NDA}
= 0.1^{+2.1}_{-0.6}.
}

\subsection{Semi-leptonic interactions} \label{sec:semilep}

The semi-leptonic CP-odd couplings $C_S^{(0)}$ and $C_S^{(1)}$ in
Eq.~(\ref{eN}) contribute to atomic and molecular EDMs by inducing
polarization of the systems. Their effects are proportional to the
nucleon-averaged coupling $C_S$ for a given nucleus:
\bea
C_S &\equiv& \frac{Z}{A} C_S^{(p)} + \frac{N}{A} C_S^{(n)}
= C_S^{(0)} + \frac{Z-N}{A} C_S^{(1)},
\eea
where $C_S^{(p)}=C_S^{(0)} + C_S^{(1)}$,
$C_S^{(n)}=C_S^{(0)} - C_S^{(1)}$, $Z$ is the proton number,
$N$ is the neutron number, and $A=Z+N$ is the mass number.

These semi-leptonic CP-odd couplings can dominantly arise from the
CP-odd electron--down-quark interaction $\mathrm{Im}(C_{eeDD})$ in
Eq.~(\ref{Xuv}):
\dis{
C_S^{(0)} = -g_S^{(0)} \mathrm{Im}\!\left(C_{eeDD}\right), \quad
C_S^{(1)} = \;\;g_S^{(1)} \mathrm{Im}\!\left(C_{eeDD}\right),
}
where $g_S^{(0)}$ and $g_S^{(1)}$ are determined by nucleon matrix
elements of light-quark bilinears,
\bea
g_S^{(0)}=\frac12 \langle N| \bar{u} u +\bar{d} d |N \rangle,\quad
g_S^{(1)}=\frac12 \langle N| \bar{u} u -\bar{d} d |N \rangle.
\eea
Since $g_S^{(1)} \sim {\cal O}(0.1)\, g_S^{(0)}$ and $(Z-N)/A \simeq -0.2$
with only small variations of order $\pm 0.03$ across paramagnetic systems
relevant for EDM experiments (see Sec.~\ref{sec:edms}), the combination
$C_S = C_S^{(0)} + \frac{Z-N}{A} C_S^{(1)}$ is nearly constant across
different atoms and molecules.

On the other hand, $C_S$ can also receive sizable contributions from
long-distance hadronic interactions induced by $\bar{g}_{0,1}$ and $d_N$,
as discussed in \cite{Flambaum:2019ejc, Mulder:2025esr, Dekens:2025skl}. We denote this
contribution as $C_S^{\rm IR}$. The result of
Ref.~\cite{Flambaum:2019ejc} can be written as
\dis{
\frac{G_F}{\sqrt{2}} C_S^{\rm IR}(\bar{g}_0,\bar{g}_1, d_p, d_n)
=
-\left(\frac{Z}{A} \xi_p + \frac{N}{A} \xi_n \right)
\frac{3\alpha m_e}{2\pi}
+
\left(\frac{Z}{A} \frac{\mu_p}{\mu_N} \frac{d_p}{e}
+ \frac{N}{A} \frac{\mu_n}{\mu_N} \frac{d_n}{e}\right)
\frac{8 \alpha^2 m_e}{m_N p_F} \ln A,
\label{CSIR}
}
where $\mu_{p,n}$ are the nucleon magnetic dipole moments, $\mu_N$ is
the nuclear magneton, and $p_F \simeq 250$ MeV is the nuclear Fermi
momentum. The coefficients $\xi_{p,n}$ are given by
\bea
\xi_p &\simeq&
-\frac{\alpha}{\pi f_\pi m_{\pi^0}^2}
\left[
(\bar{g}_1+\bar{g}_0) \ln \frac{m_{\rho}}{m_e}
+ \frac{\bar{g}_0^{\eta NN}}{\sqrt{3}}
\frac{m_{\pi^0}^2}{m_\eta^2}
\frac{f_\pi}{f_\eta}
\ln \frac{m_\rho}{m_e}
\right]
\nonumber\\
&&
-\frac{\alpha g_A}{6 f_\pi m_{\pi^\pm} m_N}
\frac{\mu_n}{\mu_N} \bar{g}_0
\left(\ln \frac{m_{\pi^\pm}}{m_e}+1.77\right),
\nonumber\\[2mm]
\xi_n &\simeq&
-\frac{\alpha}{\pi f_\pi m_{\pi^0}^2}
\left[
(\bar{g}_1-\bar{g}_0) \ln \frac{m_{\rho}}{m_e}
+ \frac{\bar{g}_0^{\eta NN}}{\sqrt{3}}
\frac{m_{\pi^0}^2}{m_\eta^2}
\frac{f_\pi}{f_\eta}
\ln \frac{m_\rho}{m_e}
\right]
\nonumber\\
&&
+\frac{\alpha g_A}{6 f_\pi m_{\pi^\pm} m_N}
\frac{\mu_p}{\mu_N} \bar{g}_0
\left(\ln \frac{m_{\pi^\pm}}{m_e}+1.77\right),
\eea
where $g_A = 1.27$ is the nucleon axial-vector coupling, and we use the improved result of Ref. \cite{Mulder:2025esr} by full two loop computations for the charged pion loop contributions involving $\ln (m_{\pi^\pm}/m_e)$.\footnote{Compared with Ref. \cite{Flambaum:2019ejc}, it has an overall factor 2/3 and the extra term +1.77 which amounts to 30\% correction to the logarithmic term.} For $C_S^{\rm IR}$ arising from the nucleon EDMs $(d_p, d_n)$ in Eq.~(\ref{CSIR}), we use the result of Ref.~\cite{Flambaum:2019ejc}, which is based on a simple non-interacting Fermi-gas model of the nucleus. We note, however, that Ref.~\cite{Dekens:2025skl} provides a more dedicated computation of $C_S^{\rm IR}(d_p, d_n)$ using nuclear matrix elements and shell-model calculations.

For heavy nuclei with $A \sim 200$ and $Z/A \simeq 0.4$,
Eq.~(\ref{CSIR}) yields
\dis{
C_S^{\rm IR}(\bar{g}_0,\bar{g}_1, d_p, d_n)
\simeq
-0.55\, \bar{g}_0
+ 2.2\, \bar{g}_1
+ 3.4\, \frac{d_p}{e\,\mathrm{fm}}
- 3.5\, \frac{d_n}{e\,\mathrm{fm}},
\label{CSIRn}
}
where the coefficients vary by less than $10\%$ across the paramagnetic
molecules used in EDM experiments.

\section{Electric dipole moments of nuclei, atoms, and molecules} \label{sec:edms}

In this section, we discuss EDMs of nuclei, atoms, and molecules
induced by the IR CP-violating (CPV) parameters introduced in the
previous section. We summarize the current theoretical computations for
selected light nuclei, atoms, and molecules whose EDMs are being
measured or targeted in ongoing and future experiments. At leading
order, the EDM $d_i$ of a system $i$ is linearly related to the IR CPV
parameters as
\dis{
d_i = \sum_j P_{ij} X_j^{\rm IR},
\label{diXj}
}
where $X_j^{\rm IR}$ is defined in Eq.~(\ref{XIR}). In the following, we
summarize the corresponding matrix elements $P_{ij}$.

\subsection{Light nuclei and diamagnetic atoms} \label{sec:dia}

The EDMs of charged light nuclei such as the proton, deuteron ($D$),
and helion ($^3\mathrm{He}^{++}$) can be directly measured in storage
ring experiments \cite{pEDM:2025nlu, CPEDM:2019nwp}. Currently, a
precursor storage-ring EDM experiment is underway at COSY by the JEDI
collaboration, targeting the deuteron EDM using a magnetic ring
\cite{Shmakova:2023yrb, Lehrach:2012eg}. Future electric and magnetic
storage-ring experiments are expected to probe the EDMs of such charged
light nuclei down to the level of $10^{-29}\, e\,\mathrm{cm}$
\cite{pEDM:2025nlu}.

Since light nuclei are relatively simple systems composed of a few
nucleons, theoretical calculations of their EDMs in terms of the IR CPV
parameters carry uncertainties at the level of $\lesssim 50\%$. In the
following, we use the results of Ref.~\cite{Bsaisou:2014zwa}:
\bea
d_D &=& 0.94(1)\,(d_n+d_p) + 0.18(2)\,\bar{g}_1 \, e \,\mathrm{fm},
\label{dD} \\
d_{\mathrm{He}} &=& 0.9\, d_n - 0.03(1)\, d_p
\nonumber\\
&&
+ \Big[
0.11(1)\,\bar{g}_0
+ 0.14(2)\,\bar{g}_1
- \left(0.04(2)\, C_1 - 0.09(2)\, C_2 \right)\mathrm{fm}^{-3}
\Big] \, e \,\mathrm{fm}.
\label{dHe}
\eea

On the other hand, strong experimental bounds already exist for neutral
diamagnetic atoms with heavy nuclei, such as $^{199}\mathrm{Hg}$,
$^{129}\mathrm{Xe}$, $^{171}\mathrm{Yb}$, and $^{225}\mathrm{Ra}$,
obtained using atomic spin-precession spectroscopy. We summarize the
current best limits on EDMs of various systems in
Table~\ref{tab:limits}. For a comprehensive overview of the experimental
status, see Ref.~\cite{Degenkolb:2024eve}.

For the matrix elements in Eq.~(\ref{diXj}), we adopt the summary given
in Table~5 of Ref.~\cite{Degenkolb:2024eve}, which is based on
Refs.~\cite{Hubert:2022pnl, Engel:2013lsa} for $^{199}$Hg,
Refs.~\cite{Hubert:2022pnl, Dmitriev:2004fk} for $^{129}$Xe,
Refs.~\cite{Dzuba:2009kn, Flambaum:2019kbn, Dzuba:2007zz} for
$^{171}$Yb, and Refs.~\cite{Chupp:2014gka, Dzuba:2009kn,
Flambaum:2019kbn} for $^{225}$Ra. In addition, we update the coefficients
of $\bar{g}_0$, $\bar{g}_1$, $C_1$, and $C_2$ for $^{225}$Ra using
Ref.~\cite{Dobaczewski:2018nim}. The resulting EDMs are
\dis{
d_{\rm Hg} = -2.26(23) \cdot 10^{-4} \left[ 0.6^{+1.33}_{-0.12} d_n + 0.06^{+0.20}_{-0.01} d_p + \frac{g_A m_N}{f_\pi} \left(0.01^{+0.04}_{-0.005} \,\bar{g}_0 +0.02^{+0.07}_{-0.05} \,\bar{g}_1 \right) e\,\textrm{fm} \right],  \label{dHg}
} 
\dis{
d_{\rm Xe} = 3.62(25) \cdot 10^{-5} \left[ 0.63^{+0.16}_{-0.12} d_n + 0.14(3) d_p + \frac{g_A m_N}{f_\pi} \left(-0.008^{+0.003}_{-0.042} \,\bar{g}_0 +0.006^{+0.044}_{-0.003} \,\bar{g}_1 \right) e\,\textrm{fm} \right],  \label{dXe} 
}
\dis{
d_{\rm Yb} =-2.10^{+0.22}_{-0.0} \cdot 10^{-4} \left[ 0.54^{+0.13}_{-0.11} d_n + 0.054^{+0.016}_{-0.014} d_p + \frac{g_A m_N}{f_\pi} \left(0.01^{+0.02}_{-0.0} \,\bar{g}_0 +0.02^{+0.034}_{-0.027} \,\bar{g}_1 \right) e\,\textrm{fm} \right],  \label{dYb}
}
\dis{ 
d_{\rm Ra} =-8.5^{+0.25}_{-0.3} \cdot 10^{-4} \bigg[ 0.63^{+0.16}_{-0.12} d_n + 0.14^{+0.04}_{-0.03} d_p&+\frac{g_A m_N}{f_\pi} \left(-0.2(6) \,\bar{g}_0 + 5(3) \,\bar{g}_1 \right) e\,\textrm{fm}  \\ 
&+m_N^3 \left( -0.01(3)C_1 + 0.03(2) C_2\right)  e\,\textrm{fm} \bigg],   \label{dRa} 
 }
where we have neglected contributions from the electron EDM $d_e$ and
the electron--nucleon coupling $C_S$, to which diamagnetic atoms are less
sensitive than paramagnetic systems (see the following subsection).
Theoretical uncertainties are sizable, particularly for contributions
from CP-odd nuclear interactions (i.e., $\bar{g}_0$, $\bar{g}_1$, $C_1$,
and $C_2$).
The EDMs of these neutral atoms are screened by atomic electrons, so
that the observable EDMs are proportional to the nuclear Schiff moments
\cite{Schiff:1963zz, Engel:2025uci}. This leads to a suppression of order
$10^{-5}$--$10^{-4}$ relative to nucleon EDMs. Nevertheless, as shown in
Table~\ref{tab:limits}, this suppression is compensated by the high
experimental sensitivity achievable with heavy atoms, owing to large
atom numbers, long spin-coherence times, and advanced comagnetometry
techniques (see, e.g., \cite{Chupp:2017rkp}).
Furthermore, systems with octupole-deformed nuclei, such as
$^{225}\mathrm{Ra}$, can exhibit a strong enhancement of Schiff moments
induced by CP-odd nuclear interactions
\cite{Auerbach:1996zd, Spevak:1996tu, Flambaum:2024xjs}. Although EDM
experiments for such systems are still at an early stage, they are
expected to provide powerful probes of CP-odd nuclear interactions in
the future, with projected sensitivities reaching the level of
$10^{-28}\, e\,\mathrm{cm}$ for $d_{\rm Ra}$
\cite{Bishof:2016uqx}.

Given the current experimental bounds summarized in
Table~\ref{tab:limits} and the theoretical expressions in
Eqs.~(\ref{dHg})--(\ref{dRa}), the EDM of $^{199}$Hg provides the most
stringent constraints on $d_p$, $\bar{g}_0$, and $\bar{g}_1$ at present.
Assuming no cancellations among the contributions of $d_p$, $d_n$,
$\bar{g}_0$, and $\bar{g}_1$ to $d_{\rm Hg}$, the experimental upper
limit on $d_{\rm Hg}$ implies~\cite{Graner:2016ses}
\dis{
|d_p| < 2.0 \times 10^{-25}\, e\,\mathrm{cm}, \quad
|\bar{g}_0| < 2.3 \times 10^{-12}, \quad
|\bar{g}_1| < 1.1 \times 10^{-12}.
\label{Hglim}
}

\begin{table}[ht]
\centering
\begin{small}
\begin{tabular}{c|c}
\toprule
System $i$&   Upper limit on $|d_i|$ $\left[\text{$e$\,cm}\right]$  \\
\midrule
$n$\, 
 & $2.2\cdot 10^{-26}$  \cite{Abel:2020pzs}  \\ [0mm]
\midrule
$^{199}$Hg
& $7.4\cdot 10^{-30}$  \cite{Graner:2016ses} \\[0mm]
$^{129}$Xe 
& $1.4\cdot 10^{-27}$  \cite{Sachdeva:2019rkt, Allmendinger:2019jrk} \\[0mm]
$^{171}$Yb
& $1.5\cdot 10^{-26}$  \cite{Zheng:2022jgr} \\[0mm]
$^{225}$Ra 
& $1.4\cdot 10^{-23}$ \cite{Bishof:2016uqx} \\[0mm]
\midrule
HfF$^+$ 
& $4.8\cdot 10^{-30}$  \cite{Roussy:2022cmp} \\[0mm]
ThO  
& $1.1\cdot 10^{-29}$ \cite{ACME:2018yjb}\\[0mm]
YbF 
& $1.2\cdot 10^{-27}$ \cite{Hudson:2011zz}\\[0mm]
\bottomrule
\end{tabular}
\end{small}
\caption{Experimental upper limits on the EDMs $|d_i|$ of system $i$ are
summarized. For polar molecules (HfF$^+$, ThO, YbF), the quoted EDM
corresponds to a quantity defined through the $P$-odd and $T$-odd
frequency shift of spin precession, rather than the permanent molecular
EDM arising from an asymmetric charge distribution within the molecule,
which exists independently of fundamental CP-violating interactions.
For a precise definition, see Sec.~\ref{sec:para}.}
\label{tab:limits}
\end{table}

\subsection{Paramagnetic systems} \label{sec:para}

Paramagnetic systems contain at least one unpaired electron in atoms
or molecules and are therefore sensitive to interactions involving
electrons. In particular, polar molecules can be highly sensitive to
CP-violating (CPV) observables such as the electron EDM $d_e$ and the
electron--nucleon coupling $C_S$\footnote{Ref. \cite{Baruch:2024frj} shows that polar molecules can also provide sensitive probes for CPV nucleon-nucleon long range interactions mediated by a new light particle whose mass is below about 10 keV which corresponds to the inter-atomic distance scale of polar molecules.}, because they can possess strong
internal effective electric fields arising from their asymmetric
molecular structure. In the following, we consider HfF$^+$, ThO, and YbF.

Since polar molecules already have nonzero permanent EDMs associated
with asymmetric charge distributions, which do not violate $P$ and $T$
symmetries, the $P$-odd and $T$-odd component induced by
fundamental CPV interactions is conventionally characterized by the
corresponding frequency shift of electron spin precession, which is
directly measurable in EDM experiments.
The $P$-odd and $T$-odd frequency shifts induced by $d_e$ and $C_S$ are
summarized in Refs.~\cite{Fleig:2018bsf, Degenkolb:2024eve}, based on
Refs.~\cite{Petrov:2007zz, Skripnikov:2017cnj, Fleig:2017mls} for HfF$^+$,
Refs.~\cite{Meyer:2008gc, Dzuba:2011avf, 10.1063/1.4968229, Denis:2016loy}
for ThO, and Refs.~\cite{Dzuba:2011avf, Abe:2014nhq, PhysRevA.93.042507}
for YbF:
\bea
\omega_{\rm HfF^+}&=&3.49(14)\times10^{28}\, d_e \,[\mathrm{mrad/s}][e\,\mathrm{cm}]^{-1}
+3.20(13)\times10^{8}\, C_S \,[\mathrm{mrad/s}],
\label{HfF}\\
\omega_{\rm ThO}&=&-1.206(49)\times10^{29}\, d_e \,[\mathrm{mrad/s}][e\,\mathrm{cm}]^{-1}
-1.816(73)\times10^{9}\, C_S \,[\mathrm{mrad/s}],
\label{ThO}\\
\omega_{\rm YbF}&=&-1.96(15)\times10^{28}\, d_e \,[\mathrm{mrad/s}][e\,\mathrm{cm}]^{-1}
-1.76(20)\times10^{8}\, C_S \,[\mathrm{mrad/s}],
\label{YbF}
\eea
with theoretical uncertainties typically below the $10\%$ level.
In these expressions, the coefficient multiplying $d_e$ corresponds to
the effective electric field $E_{\rm eff}$ intrinsic to the molecular
structure, independent of any applied external electric field.\footnote{
An external electric field is nevertheless required in experiments to
polarize the molecule and align the internal effective field.}
Following Ref.~\cite{Degenkolb:2024eve}, one may define an effective
EDM for a given molecule $i$ as $d_i \equiv \omega_i/E_{\rm eff}$; the
corresponding experimental upper limits on $|d_i|$ are given in
Table~\ref{tab:limits}.

Paramagnetic atoms such as Tl and Cs are also sensitive to $d_e$ and
$C_S$, although their sensitivities are typically two to three orders
of magnitude weaker than those of polar molecules, primarily due to the
absence of strong internal effective electric fields in atoms (see,
e.g., \cite{Pospelov:2005pr, Hudson:2002az}). On the other hand, such
systems can provide sensitive probes of nuclear spin-dependent CPV
electron--nucleon interactions of the form
$C_P \bar{e} e \bar{N} i\gamma_5 N
+ C_T \bar{e} \sigma^{\mu \nu} e \bar{N} \sigma_{\mu \nu} i \gamma_5 N$
\cite{Ginges:2003qt, Degenkolb:2024eve}. In this work, however, we do not
consider these contributions, as the corresponding nuclear spin-dependent
interactions are negligible for the UV sources discussed in
Sec.~\ref{sec:UVsrc}.


Currently, the most stringent constraints on $d_e$ and $C_S$ are obtained
from the EDM measurement of HfF$^+$, yielding~\cite{Roussy:2022cmp}
\dis{
|d_e| < 4.1\times 10^{-30}\, e\,\mathrm{cm}, \quad
|C_S| < 4.4\times 10^{-10},
\label{HfFlim}
}
at 90\% CL, assuming no cancellation between the contributions of $d_e$
and $C_S$ to $\omega_{\mathrm{HfF}^+}$.

Paramagnetic systems can also probe hadronic IR CPV parameters
($d_p$, $d_n$, $\bar{g}_0$, $\bar{g}_1$) through the induced coupling
$C_S^{\rm IR}$ in Eq.~(\ref{CSIR}), which arises from long-distance
hadronic interactions. To estimate the relative sensitivity of polar
molecules to these hadronic CPV parameters, we parametrize the
$P$-odd and $T$-odd frequency shift of a polar molecule $X$ as
\dis{
\frac{\omega_X}{0.01\,\mathrm{mrad/s}} =
\alpha_X \frac{d_e}{10^{-30}\,[e\,\mathrm{cm}]}
+ \beta_X \frac{C_S}{10^{-10}},
}
where $\alpha_X$ and $\beta_X$ are typically of
$\mathcal{O}(1)$, except for ThO, for which
$\alpha_{\mathrm{ThO}} \sim \beta_{\mathrm{ThO}} \sim \mathcal{O}(10)$.
The normalization factors are chosen to be close to the current
experimental limits in Eq.~(\ref{HfFlim}). Using
Eq.~(\ref{CSIRn}), we obtain
\dis{
\frac{C_S^{\rm IR}}{10^{-10}} \simeq
-0.55 \frac{\bar{g}_0}{10^{-10}}
+ 2.2 \frac{\bar{g}_1}{10^{-10}}
+ 3.4 \frac{d_p}{10^{-23}\,[e\,\mathrm{cm}]}
- 3.5 \frac{d_n}{10^{-23}\,[e\,\mathrm{cm}]}.
}
The coefficients vary by less than $10\%$ across the polar molecules
considered in EDM experiments. Comparing with Eq.~(\ref{Hglim}), one
finds that the current sensitivity of polar molecules to hadronic IR CPV
parameters is approximately two to three orders of magnitude weaker than
that of diamagnetic atoms.


\section{Identification of the UV sources with EDM data} \label{sec:iden}

We now investigate \emph{the EDM inverse problem} based on the
state-of-the-art computations summarized in
Secs.~\ref{sec:IRsrc} and \ref{sec:edms}. Our goal is to assess whether
the six classes of UV CP-violating (CPV) sources in
Eq.~(\ref{Xuv}) can be distinguished using future EDM data, taking into
account theoretical uncertainties.
Among the sources in Eq.~(\ref{Xuv}), the hadronic CPV sources
$X_i^{\rm UV} \in \left\{ \bar{\theta}, w, \tilde{d}_q, d_q \right\}$
are most sensitively probed by light nuclei and diamagnetic atoms, as
discussed in Sec.~\ref{sec:dia}, whereas the (semi-)leptonic sources
$X_i^{\rm UV} \in \left\{ d_e, \mathrm{Im}(C_{eeDD}) \right\}$ are best
probed by polar molecules, as discussed in Sec.~\ref{sec:para}.
We therefore analyze these two classes separately.

\subsection{Hadronic sources}

\subsubsection{Single source}

A simple and plausible working assumption is that CP violation is
dominated by a single hadronic UV source among
$X_i^{\rm UV} \in \left\{ \bar{\theta}, w, w^{\rm PQ},
\tilde{d}_q, \tilde{d}_q^{\rm PQ}, d_q \right\}$, where the superscript
``PQ'' denotes quantities in the presence of the PQ mechanism.
Under this assumption, one can predict characteristic ratios of EDMs
for light nuclei and diamagnetic atoms. From
Eqs.~(\ref{dD})--(\ref{dRa}), we obtain
\bea
\frac{d_D}{d_n} &=&
0.94(1)\left(1+\frac{d_p}{d_n}\right)
+ 1.05(12)\,\frac{e\bar{g}_1}{\Lambda_\chi d_n},
\label{Dn} \\
\frac{d_{\rm He}}{d_n} &=&
0.9 - 0.03(1)\frac{d_p}{d_n}
+ 0.64(6)\,\frac{e\bar{g}_0}{\Lambda_\chi d_n}
+ 0.82(12)\,\frac{e\bar{g}_1}{\Lambda_\chi d_n}
- 0.18(9)\,\frac{e f_\pi^2 C_1}{d_n},
\label{Hen} \\
\frac{d_{\rm Hg}}{d_n} &=&
\left(
-2.7(17)
- 0.35(24)\frac{d_p}{d_n}
- 5(4)\,\frac{e\bar{g}_0}{\Lambda_\chi d_n}
- 5(11)\,\frac{e\bar{g}_1}{\Lambda_\chi d_n}
\right)\times 10^{-4},
\label{Hgn} \\
\frac{d_{\rm Xe}}{d_n} &=&
\left(
2.4(5)
+ 0.51(11)\frac{d_p}{d_n}
- 8(6)\,\frac{e\bar{g}_0}{\Lambda_\chi d_n}
+ 7(7)\,\frac{e\bar{g}_1}{\Lambda_\chi d_n}
\right)\times 10^{-5},
\label{Xen} \\
\frac{d_{\rm Yb}}{d_n} &=&
\left(
-4(5)
- 0.109(30)\frac{d_p}{d_n}
- 3.1(16)\,\frac{e\bar{g}_0}{\Lambda_\chi d_n}
- 4(5)\,\frac{e\bar{g}_1}{\Lambda_\chi d_n}
\right)\times 10^{-4},
\label{Ybn} \\
\frac{d_{\rm Ra}}{d_n} &=&
\left(
-5.5(12)
- 1.24(30)\frac{d_p}{d_n}
\right)\times 10^{-4}
+ 0.01(4)\,\frac{e\bar{g}_0}{\Lambda_\chi d_n}
- 0.33(20)\,\frac{e\bar{g}_1}{\Lambda_\chi d_n}
\nonumber\\
&&
+ 0.004(13)\,\frac{e f_\pi^2 C_1}{d_n},
\label{Ran}
\eea
where $\Lambda_\chi = 4\pi f_\pi$.
These expressions show that each of the ratios
$d_p/d_n$, $e\bar{g}_0/(\Lambda_\chi d_n)$,
$e\bar{g}_1/(\Lambda_\chi d_n)$, and
$e f_\pi^2 C_1/d_n$ contributes comparably to the EDM ratios
$d_X/d_n$ ($X = D$, He, Hg, Xe, Yb, Ra), unless the corresponding
contribution is negligibly small.

Table~\ref{tab:ratios1} summarizes the relative importance of these
contributions for different dominant CPV sources, based on the discussion
in Sec.~\ref{sec:IRsrc}. In particular, if CP violation is dominated by
quark CEDMs ($\tilde{d}_q$ or $\tilde{d}_q^{\rm PQ}$), the ratios
$d_X/d_n$ are largely controlled by
$e\bar{g}_1/(\Lambda_\chi d_n)$. In contrast, if CP violation is dominated
by quark EDMs ($d_q$), the ratios are primarily determined by $d_p/d_n$.

\begin{table}[ht]
\begin{center}
\begin{tabular}{  C{2cm} | C{2.5cm}  C{3.8cm}  C{3.8cm}  }
\toprule
                    & $\bar{\theta}$ & $w$ & $w^{\rm PQ}$\\ \midrule
$\dfrac{d_p}{d_n}$  &    $-1.5$ &  $-0.90(3)$ &    $  -0.90(3) $      \\  [4mm]
$\dfrac{e\bar{g}_0}{\Lambda_\chi d_n}$  
  & 3.5(6)     
  & $ 0.08(28) r(\Lambda) \pm \frac{m_u+m_d}{f_\pi}$   
  & $ 0.7(5) r(\Lambda) \pm \frac{m_u+m_d}{f_\pi} $ \\ [5mm]
$\dfrac{e \bar{g}_1}{\Lambda_\chi d_n}$  
  & $-0.8(5)$      
  & $ -4.8(33) r(\Lambda) \pm 0.11(9)$     
  & $ -4.8(33) r(\Lambda) \pm 0.11(9)$   \\ [5mm]
$\dfrac{e f_\pi^2 C_1}{d_n}$  
  & -    
  & $0.1^{+2.1}_{-0.6}$     
  & $ 0.1^{+2.1}_{-0.6}$    \\ \bottomrule
\end{tabular}
\end{center}

\begin{center}
\begin{tabular}{  C{2cm} | C{2.5cm}  C{3.8cm}  C{3.8cm}   }
\toprule
                       & $d_q$ & $ \tilde{d}_q $  & $\tilde{d}_q^{\rm PQ}$\\ \midrule
$\dfrac{d_p}{d_n}$  
  & $ -\dfrac{4d_u-d_d}{d_u-4d_d} $     
  & $ \dfrac{0.81(6) \tilde{d}_u -1.51(4) \tilde{d}_d }{ 0.64(4) \tilde{d}_u -0.18(4) \tilde{d}_d} $  
  & $- \dfrac{8 \tilde{d}_u +  \tilde{d}_d }{2 \tilde{d}_u +4\tilde{d}_d} $ \\ [4mm]
$\dfrac{e\bar{g}_0}{\Lambda_\chi d_n}$  
  & -     
  & $\dfrac{4(11) \tilde{d}_u + 2(10) \tilde{d}_d}{4.4(8) \tilde{d}_u - 1.2(11)\tilde{d}_d}$  
  & $ \dfrac{5.6(22) (\tilde{d}_u + \tilde{d}_d)}{ \tilde{d}_u +2 \tilde{d}_d}$   \\ [5mm]
$\dfrac{e \bar{g}_1}{\Lambda_\chi d_n}$  
  & -          
  & $\dfrac{-483(165) (\tilde{d}_u - \tilde{d}_d)}{4.4(8) \tilde{d}_u - 1.2(11)\tilde{d}_d}$  
  & $ \dfrac{99(37) (\tilde{d}_u - \tilde{d}_d)}{ \tilde{d}_u +2 \tilde{d}_d}$   \\ [5mm]
$\dfrac{e f_\pi^2 C_1}{d_n}$  
  & -          
  & -  
  & -   \\ \bottomrule
\end{tabular}
\end{center}
\caption{Ratios of IR CPV parameters to the neutron EDM arising from a single dominant UV CPV operator. For $d_p/d_n$, the ratios have to be understood to have at least 10\% error from higher-order contributions in the OPE of the QCD sum rules for nucleon EDMs.}
\label{tab:ratios1}
\end{table}

\begin{table}[ht]
\begin{adjustwidth}{-2cm}{0cm}
\begin{center}
\begin{tabular}{  c | c  c  c  c  c  }
\toprule
                    & $\bar{\theta}$ & $w,\, w^{\rm PQ}$ & $d_q$ & $ \tilde{d}_q $  & $\tilde{d}_q^{\rm PQ} $\\ \midrule
$\dfrac{d_p}{d_n}$  &   $ -1.5 $  &  -0.90(3) &    $ -\frac{4d_u-d_d}{d_u-4d_d} $     &      $ \frac{0.81(6) \tilde{d}_u -1.51(4) \tilde{d}_d }{ 0.64(4) \tilde{d}_u -0.18(4) \tilde{d}_d} $  &   $ \frac{8 \tilde{d}_u +  \tilde{d}_d }{2 \tilde{d}_u +4\tilde{d}_d} $     \\ [5mm]
$\dfrac{d_D}{d_n}$  &       -1.3(5)     &    $ -5.1(35) r(\Lambda) +0.09(21)$   &          $ \frac{ -2.82(3)(d_u+d_d)}{d_u-4d_d} $      &    $\frac{-505(183) (\tilde{d}_u - \tilde{d}_d)}{4.4(8) \tilde{d}_u - 1.2(11)\tilde{d}_d} $  &  $ \frac{105(41) (\tilde{d}_u - \tilde{d}_d) }{\tilde{d}_u +2\tilde{d}_d} $   \\ [5mm]
$\dfrac{d_{\rm He}}{d_n}$  &    $ 2.5(6) $      &   $-3.9(28) r(\Lambda) +0.77(31) $     &          $ 0.9 + 0.03(1) \frac{4 d_u -d_d}{d_u-4d_d}  $            &  $\frac{-396(147) (\tilde{d}_u - \tilde{d}_d)}{4.4(8) \tilde{d}_u - 1.2(11)\tilde{d}_d} $   &   $ \frac{81(32) (\tilde{d}_u -\tilde{d}_d) }{\tilde{d}_u +2\tilde{d}_d} $   \\ [5mm]
$10^4 \dfrac{d_{\rm Hg}}{d_n}$  &    $ -15(17) $      &   $25(53) r(\Lambda) -2.4(20) $     &          $  \frac{-1.3(19) d_u +11(7)d_d}{d_u-4d_d}  $            &  $\frac{3(5)\times10^3(\tilde{d}_u - \tilde{d}_d)}{4.4(8) \tilde{d}_u - 1.2(11)\tilde{d}_d} $   &   $ \frac{-0.5(11) \times 10^3 (\tilde{d}_u -\tilde{d}_d) }{\tilde{d}_u +2\tilde{d}_d} $   \\  [5mm]
$10^5 \dfrac{d_{\rm Xe}}{d_n}$  &    $ 31(24) $      &   $-36(40) r(\Lambda) +1.9(17) $     &          $ \frac{0.3(7) d_u -8.9(21)d_d}{d_u-4d_d}   $            &  $\frac{-3.6(34) \times 10^3 (\tilde{d}_u - \tilde{d}_d)}{4.4(8) \tilde{d}_u - 1.2(11)\tilde{d}_d} $   &   $ \frac{0.74(71) \times 10^3 (\tilde{d}_u -\tilde{d}_d) }{\tilde{d}_u +2\tilde{d}_d}$  \\ [5mm]
$10^4 \dfrac{d_{\rm Yb}}{d_n}$  &    $ -9(7) $      &   $17(26) r(\Lambda) -1.0(8) $     &          $  \frac{-0.66(27) d_u +4.3(10)d_d}{d_u-4d_d}  $            &  $\frac{1.7(24)\times10^3(\tilde{d}_u - \tilde{d}_d)}{4.4(8) \tilde{d}_u - 1.2(11)\tilde{d}_d} $   &   $ \frac{-0.36(49) \times 10^3 (\tilde{d}_u -\tilde{d}_d) }{\tilde{d}_u +2\tilde{d}_d} $   \\  [5mm]
$\dfrac{d_{\rm Ra}}{d_n}$  &    $ 0.31(27) $      &   $1.6(14) r(\Lambda) +0.00(7) $     &      $ \frac{-0.6(17) d_u +21(5)d_d}{d_u-4d_d} \cdot 10^{-4}$         &  $\frac{160(110) (\tilde{d}_u - \tilde{d}_d)}{4.4(8) \tilde{d}_u - 1.2(11)\tilde{d}_d} $   &   $ \frac{-33(23) (\tilde{d}_u -\tilde{d}_d) }{\tilde{d}_u +2\tilde{d}_d} $   
  \\ \bottomrule
\end{tabular} 
\end{center}
\end{adjustwidth}
\caption{Predicted nuclear and atomic EDMs, normalized to the neutron EDM, for scenarios with a single dominant UV CPV operator. For $d_p/d_n$ and $d_D/d_n$, the ratios have to be understood to have at least 10\% error from higher-order contributions in the OPE of the QCD sum rules for nucleon EDMs.}
\label{tab:ratios2}
\end{table}

\begin{figure}[h]
\begin{center}
 \begin{tabular}{l}
 \hspace{-1.3cm}
    \includegraphics[scale=0.35]{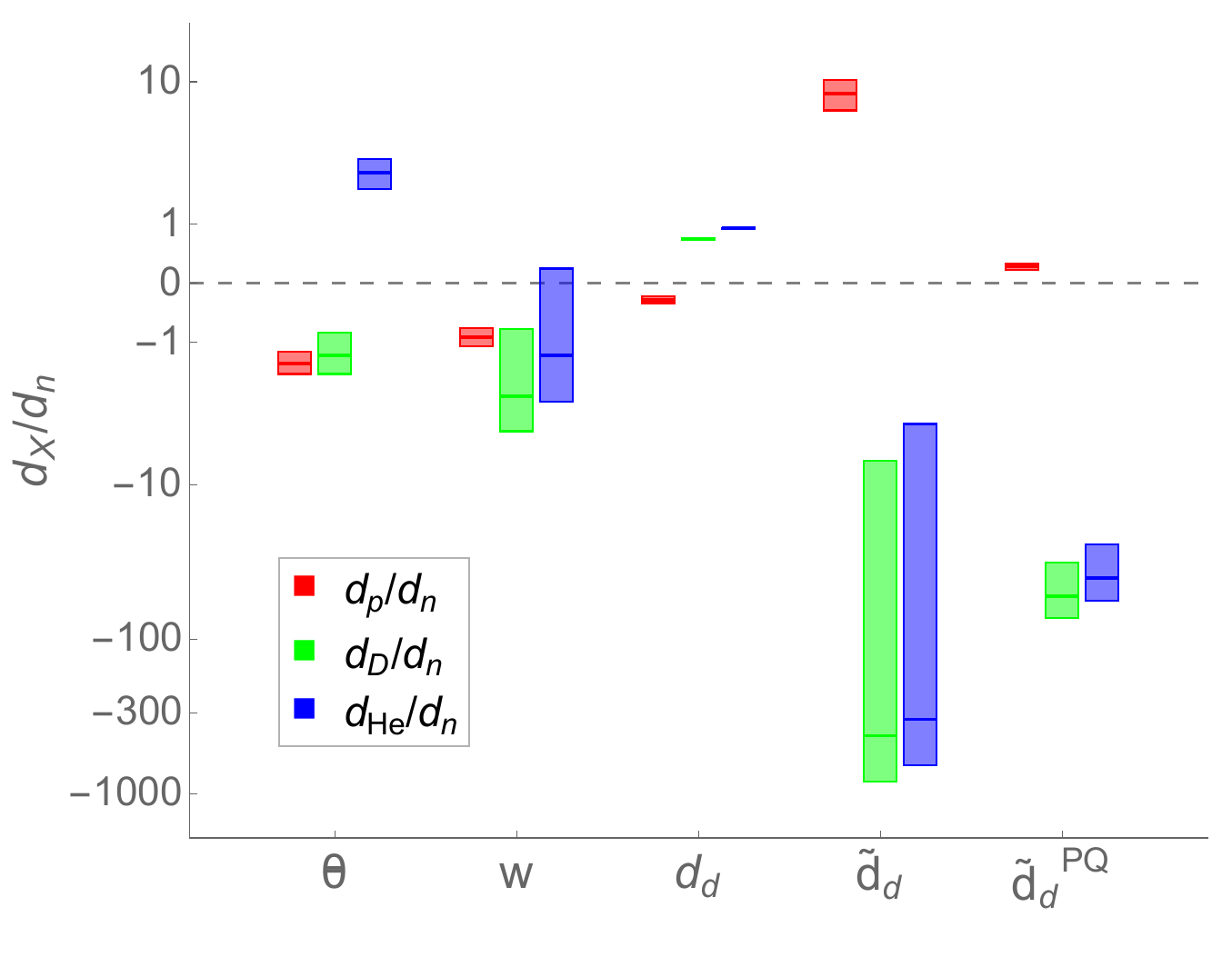}
    \hspace{0.0cm}
    \includegraphics[scale=0.35]{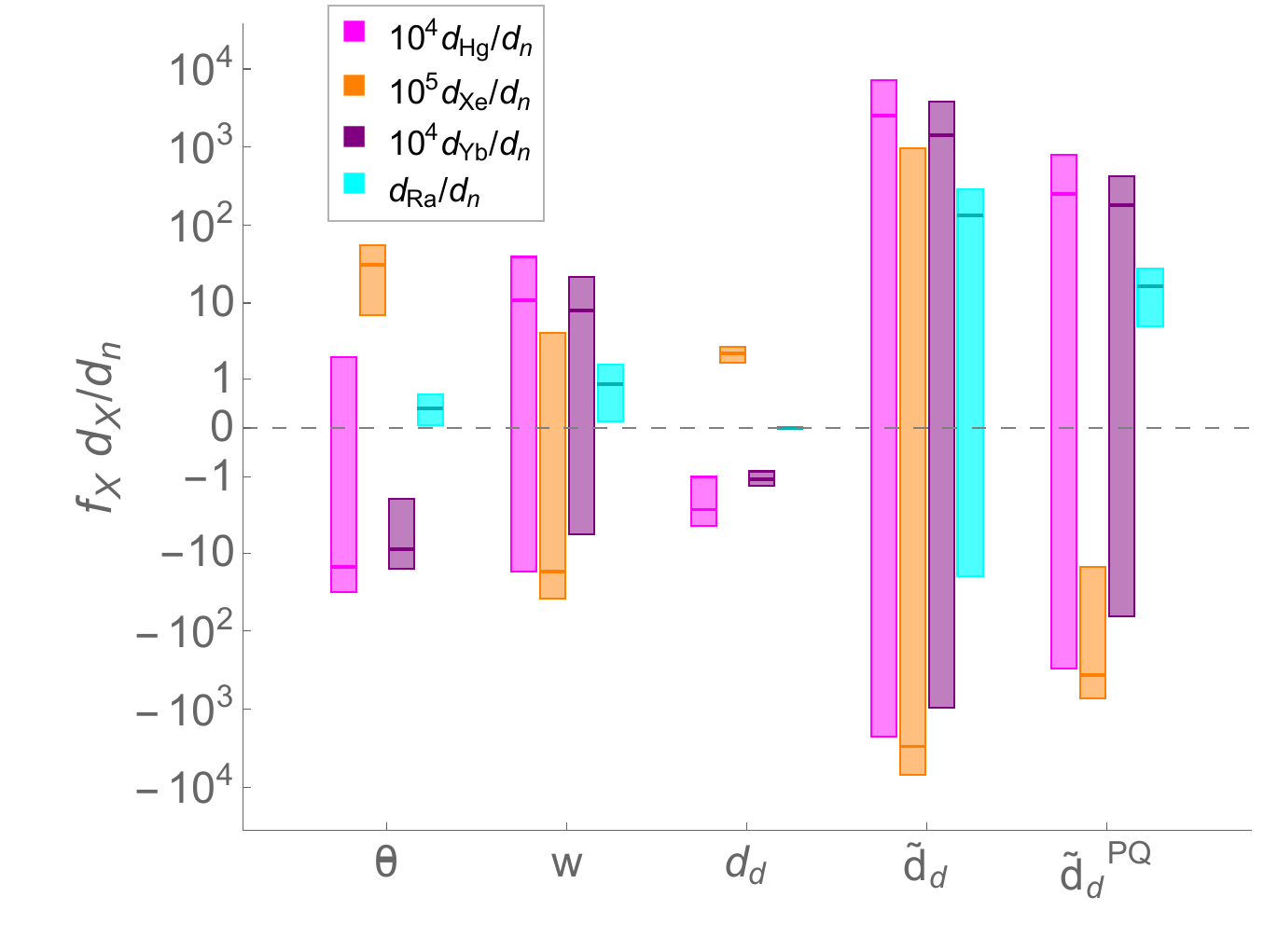} 
   \end{tabular}
  \end{center}
  \caption{
 Illustration of Table~\ref{tab:ratios2} for a specific scenario with
$\Lambda = 10$ TeV in the gluon CEDM--dominated case and
down-quark (C)EDM dominance over the up-quark (C)EDM.
In the right panel, $f_X = 10^4$ ($X = \mathrm{Hg}, \mathrm{Xe}$),
$10^5$ ($X = \mathrm{Yb}$), and $1$ ($X = \mathrm{Ra}$) denote the
normalization factors accounting for Schiff screening and octupole
enhancement.}
\label{fig:ratios2}
\end{figure}

Table~\ref{tab:ratios2} summarizes the predicted EDM ratios for
different dominant sources of CP violation, under the assumption that a
single class of operators dominates. In each case, the ratios depend on
at most one parameter: $r(\Lambda)$ for the $w$-dominated scenario,
$d_u/d_d$ for the $d_q$-dominated scenario, and
$\tilde{d}_u/\tilde{d}_d$ for the $\tilde{d}_q$-dominated scenario.
Therefore, in principle, measurements of two independent EDM ratios in
Table~\ref{tab:ratios2} can be used to identify the underlying source of
CP violation.

The predicted values are subject to sizable theoretical uncertainties,
particularly for heavy atoms such as $^{199}$Hg, $^{225}$Ra, and
$^{129}$Xe, which can exceed $100\%$ in some cases. By contrast, the
uncertainties are relatively smaller, typically below $50\%$, for light
nuclei such as $p$, $D$, and $^3$He$^{++}$. Nevertheless, measurements
of EDM ratios can still provide valuable guidance on the nature of the
underlying CPV source.

In Fig.~\ref{fig:ratios2}, we illustrate the results of
Table~\ref{tab:ratios2} for a representative scenario in order to
provide a qualitative picture of the EDM ratios. For this plot, we
assume $\Lambda = 10$ TeV for the gluon CEDM–dominated case, and that
the down-quark (C)EDM dominates over the up-quark (C)EDM, as motivated
in supersymmetric extensions of the SM or type-II two-Higgs-doublet
models with large $\tan\beta \gg 1$.

We now highlight several notable features of the predicted EDM ratios in
Table~\ref{tab:ratios2} and Fig.~\ref{fig:ratios2}. The results indicate
that the $\tilde{d}_q$-dominated scenario, with or without the PQ
mechanism, can be clearly distinguished from the other scenarios by
large ratios of ${\cal O}(10^2)$–${\cal O}(10^3)$, driven by the sizable
CP-odd pion--nucleon coupling $\bar{g}_1$. By contrast, if the measured
ratios are of ${\cal O}(1)$, this would suggest that the dominant CPV
source is $\bar{\theta}$, $w$, or $d_q$.

For the $d_q$-dominated scenario, the predicted ratios exhibit relatively
small theoretical uncertainties, enabling a clear experimental
identification of the source. In contrast, it may be difficult to
distinguish the $\bar{\theta}$-dominated scenario from the
$w$-dominated scenario given the current theoretical uncertainties. In
this case, additional information such as relative sign differences
among certain EDM ratios (e.g., $d_D/d_n$ and $d_{\rm He}/d_n$) may be
required.

\subsubsection{Multiple sources}

The EDM ratios can deviate significantly from those in
Table~\ref{tab:ratios2} if multiple CPV operators contribute
comparably to the nuclear EDMs. In such cases, one must solve the
corresponding linear system to determine the underlying CPV sources.
Since the contribution of $C_1(w)$ is relatively small, we neglect it
as an approximation. The remaining four IR CPV parameters,
$d_n$, $d_p$, $\bar{g}_0$, and $\bar{g}_1$, can in principle be
determined from four independent EDM measurements.

Given the theoretical uncertainties in
Eqs.~(\ref{dD})--(\ref{dRa}), EDM data from light nuclei
($n$, $p$, $D$, $^3\mathrm{He}^{++}$) provide the most precise
determination. In particular, inverting Eqs.~(\ref{dD}) and
(\ref{dHe}) yields
\dis{
\begin{pmatrix}
d_n \,[\mathrm{GeV}/e]\\
d_p \,[\mathrm{GeV}/e]\\
\bar{g}_0\\
\bar{g}_1
\end{pmatrix}
=
{\cal M}_{\rm IR}
\begin{pmatrix}
d_n\\
d_p\\
d_D\\
d_{\mathrm{He}}
\end{pmatrix}
\,[\mathrm{GeV}/e],
}
where
\dis{
{\cal M}_{\rm IR} =
\begin{pmatrix}
1 & 0 & 0 & 0 \\
0 & 1 & 0 & 0 \\
-0.30(26) & 1.37(27) & -1.40(28) & 1.80(33) \\
-1.03(18) & -1.03(18) & 1.10(19) & 0
\end{pmatrix}.
}
This matrix determines the IR CPV parameters from the EDM data.

By contrast, the IR CPV parameters are poorly constrained by EDMs of
heavy diamagnetic atoms due to large theoretical uncertainties. For
example, using $^{199}\mathrm{Hg}$, $^{129}\mathrm{Xe}$,
$^{225}\mathrm{Ra}$, together with the neutron EDM, one finds
\dis{
\begin{pmatrix}
d_n \,[\mathrm{GeV}/e]\\
d_p \,[\mathrm{GeV}/e]\\
\bar{g}_0\\
\bar{g}_1
\end{pmatrix}
=
{\cal M}_{\rm IR}
\begin{pmatrix}
d_n\\
d_{\mathrm{Hg}}\\
d_{\mathrm{Xe}}\\
d_{\mathrm{Ra}}
\end{pmatrix}
\,[\mathrm{GeV}/e],
}
with
\dis{
{\cal M}_{\rm IR} =
\begin{pmatrix}
1 & 0 & 0 & 0 \\
-6(7) & -1.4(20)\times10^4 & 1.0(13)\times10^5 & 45(69) \\
-0.13(34) & -1.1(12)\times10^3 & -8(10)\times10^3 & 0(4) \\
0.00(4) & -39(142) & -0.4(10)\times10^3 & -4(4)
\end{pmatrix},
}
where most entries carry theoretical uncertainties exceeding $100\%$.
Therefore, precise EDM measurements of light nuclei are essential for
disentangling CPV sources when multiple contributions are present,
unless theoretical calculations for heavy diamagnetic atoms are
significantly improved.

After determining $(d_n, d_p, \bar{g}_0, \bar{g}_1)$, one can at most
disentangle four independent UV CPV sources among the six possible
sources $(\bar{\theta}, w, d_u, d_d, \tilde{d}_u, \tilde{d}_d)$, either
with or without the PQ mechanism. Since the presence of quark CEDMs is
signaled by a large ratio $e\bar{g}_1/(\Lambda_\chi d_n) \gg 1$, it is
useful to classify possible scenarios according to the size of this
ratio.

\begin{itemize}

\item{$e\bar{g}_1/(\Lambda_\chi d_n) \gg 1$}

A large value of this ratio indicates that quark CEDMs are likely to be
involved in the CPV source, unless there is a fine-tuned cancellation
among different contributions. In many UV scenarios that generate quark
CEDMs, such as supersymmetric extensions of the SM or type-II 2HDMs,
quark EDMs are typically of comparable magnitude, since both arise from
loops involving colored particles carrying electric charge
\cite{Chupp:2017rkp, Jung:2013hka}. By contrast, the QCD
$\theta$ term and the gluon CEDM usually arise from distinct
physical origins. It is therefore well motivated to consider scenarios
in which quark CEDMs and quark EDMs contribute simultaneously to nuclear
EDMs:
\dis{
\begin{pmatrix}
\tilde{d}_u \\
\tilde{d}_d \\
d_u \\
d_d
\end{pmatrix}
[\mathrm{GeV}]
=
{\cal M}_{\rm UV}^{(\rm PQ)}
\begin{pmatrix}
d_n \,[\mathrm{GeV}/e]\\
d_p \,[\mathrm{GeV}/e]\\
\bar{g}_0 \\
\bar{g}_1
\end{pmatrix},
}
where ${\cal M}_{\rm UV}^{(\rm PQ)}$ determines the UV parameters from
the IR ones, and depends on whether the PQ mechanism is present.
Without the PQ mechanism, we obtain
\dis{
{\cal M}_{\rm UV} =
\begin{pmatrix}
0 & 0 & -2(9) & 0.00(6) \\
0 & 0 & -2(9) & -0.02(10) \\
0.3(22) & 1(9) & 0.7(33) & 0.02(12) \\
1(9) & 0.3(22) & -0.3(17) & 0.01(6)
\end{pmatrix},
}
whose precision is rather poor, mainly due to the large theoretical
uncertainty in $\bar{g}_0(\tilde{d}_q)$ in
Eq.~(\ref{gbar0_dqt'}). In contrast, in the presence of the PQ mechanism,
we find
\dis{
{\cal M}^{\rm PQ}_{\rm UV} =
\begin{pmatrix}
0 & 0 & 0.23(8) & 0.013(4) \\
0 & 0 & 0.23(8) & -0.013(4) \\
0.327(27) & 1.31(11) & 0.38(14) & 0.022(8) \\
1.31(11) & 0.327(27) & -0.19(7) & 0.011(4)
\end{pmatrix},
}
which exhibits significantly improved precision.

\item{$e\bar{g}_1/(\Lambda_\chi d_n) \sim {\cal O}(1)$}

In this case, CPV sources other than quark CEDMs (i.e.
$\bar{\theta}$, $w$, $d_u$, and $d_d$) may contribute comparably to the
nuclear EDMs. One can then write
\dis{
\begin{pmatrix}
\bar{\theta} \\
w \,[\mathrm{GeV}]^2 \\
d_u \,[\mathrm{GeV}] \\
d_d \,[\mathrm{GeV}]
\end{pmatrix}
=
{\cal M}_{\rm UV}(\Lambda)
\begin{pmatrix}
d_n \,[\mathrm{GeV}/e]\\
d_p \,[\mathrm{GeV}/e]\\
\bar{g}_0 \\
\bar{g}_1
\end{pmatrix},
}
where ${\cal M}_{\rm UV}(\Lambda)$ depends on the scale $\Lambda$ at
which the Weinberg operator is generated, through the RG factor
$r(\Lambda)$ in Eq.~(\ref{rgdqtw}). For $\Lambda \gtrsim 1$ TeV, we find
\dis{
{\cal M}_{\rm UV}(\Lambda) =
\begin{pmatrix}
0 & 0 & 66(36) & A \pm 5.4(24)\,r^{-1} \\
0 & 0 & -2.0(17)\,r^{-1} & -9(4)\,r^{-1} \\
0.33(18) & 1.3(7) & 0.42(23) - 0.034(35)\,r^{-1} & -0.16(12)\,r^{-1} + 0.06(4) \\
1.3(7) & 0.33(18) & -0.21(12) + 0.04(4)\,r^{-1} & 0.19(14)\,r^{-1} - 0.032(18)
\end{pmatrix},
}
where the coefficient $A$ depends on the presence of the PQ mechanism:
$A = -1.0(29)$ without PQ and $A = 10(6)$ with PQ.

\end{itemize}


    


\subsection{(Semi-)leptonic sources}

Among the UV CPV sources in Eqs.~(\ref{L1})--(\ref{L3}), those involving
electrons, namely the electron EDM and the CPV electron-down quark coupling, can be probed with high sensitivity by paramagnetic systems,
in particular polar molecules, as discussed in Sec.~\ref{sec:para}.

\begin{table}[ht]
\begin{adjustwidth}{-0cm}{0cm}
\begin{center}
\begin{tabular}{  c | c  c  }
\toprule
                    & $d_e$ & $C_S$ \\ \midrule
$\dfrac{\omega_{\rm ThO}}{\omega_{\textrm{HfF}^+}}$  &    -3.46(20)  &  -5.68(32)       \\ [5mm]
$\dfrac{\omega_{\rm YbF}}{\omega_{\textrm{HfF}^+}}$  &       -0.56(5)     &    -0.55(7)   
  \\ \bottomrule
\end{tabular} 
\end{center}
\end{adjustwidth}
\caption{Predicted EDMs of polar molecules, normalized to that of HfF$^+$, for scenarios in which either $d_e$ or $C_S$ dominates.}
\label{tab:ratios3}
\end{table}

In Table~\ref{tab:ratios3}, we present the predicted ratios of the
$P$-odd and $T$-odd frequency shifts for the polar molecules HfF$^+$,
ThO, and YbF, assuming that CP violation is dominated by either $d_e$
or $C_S$. The ratios are obtained from
Eqs.~(\ref{HfF})--(\ref{YbF}). They show that $d_e$ can be clearly
distinguished from $C_S$ using the EDM data of ThO together with that of
another polar molecule as also discussed in \cite{Fleig:2018bsf}, since the theoretical uncertainties are
typically below $\sim 10\%$ level.

If the dominant CPV source is $C_S$, the relevant UV sources among those
in Eqs.~(\ref{L1})--(\ref{L3}) are the electron-down quark CPV coupling
and hadronic operators that induce $C_S$ via long-distance
electromagnetic interactions, as discussed in
Sec.~\ref{sec:semilep}. These two classes can be discriminated by EDM
measurements of light nuclei and diamagnetic atoms, since hadronic
operators generically induce sizable EDMs in those systems, whereas the
electron-down quark CPV coupling does not.

If $C_S$ originates from hadronic operators, one expects observable
signals in light nuclei or diamagnetic atoms with magnitudes given by
either $d_{n,p} \sim 10^{-13}\, e\,\mathrm{cm} \times C_S$ or
$\bar{g}_{0,1} \sim C_S$, as inferred from Eq.~(\ref{CSIRn}).

If the measured ratios deviate from the predictions in
Table~\ref{tab:ratios3}, this indicates that both $d_e$ and $C_S$
contribute comparably to the EDMs of polar molecules.\footnote{According to Ref. \cite{Baruch:2024frj}, there is another possibility that the ratios are due to CPV nucleon-nucleon long range interactions mediated by a new particle lighter than 10 keV.} In that case, one
must solve the linear system in
Eqs.~(\ref{HfF})--(\ref{YbF}) to extract $d_e$ and $C_S$ from the data.
For example, inverting Eqs.~(\ref{HfF}) and (\ref{ThO}) yields
\dis{
\begin{pmatrix}
d_e \,[e\,\mathrm{cm}]^{-1} \\
C_S
\end{pmatrix}
=
\begin{pmatrix}
7.3(13)\times 10^{-29}  &  1.29(23)\times 10^{-29} \\
-4.9(9)\times 10^{-9} &  -1.41(25)\times 10^{-9}
\end{pmatrix}
\begin{pmatrix}
\omega_{\rm HfF^+} \\
\omega_{\rm ThO}
\end{pmatrix}
[\mathrm{mrad/s}]^{-1},
\label{deCs}
}
where the theoretical uncertainties are below $20\%$. Thus, measurements
of two different polar molecules allow a determination of both $d_e$
and $C_S$.

\section{Conclusions} \label{sec:conc}

In this work, we investigate the feasibility of inferring ultraviolet (UV) sources of CP violation (CPV) from precision measurements of electric dipole moments (EDMs) of light nuclei, atoms, and molecules. We first identify six important classes of CPV operators at the QCD scale that arise in the Standard Model (SM) and in well-motivated beyond-the-SM (BSM) scenarios. These operators, listed in Eq.~(\ref{Xuv}), are the QCD $\theta$ term, the gluon chromo-electric dipole moment (CEDM), often referred to as the Weinberg three-gluon operator, quark CEDMs, quark EDMs, the electron EDM, and CP-odd electron--down-quark interactions.

Among these six classes, the four hadronic operators that do not involve electrons can be probed with high sensitivity through EDM measurements of light nuclei and diamagnetic atoms. If the observed EDMs are dominated by a single operator class, the resulting EDM ratios depend on only one parameter. Consequently, precision measurements of at least three different nuclei or atoms, providing two independent EDM ratios, can in principle identify the dominant hadronic UV CPV source. Given current theoretical uncertainties, we find that EDM measurements of three light nuclei, such as the neutron ($n$), proton ($p$), and deuteron ($D$), are particularly powerful for discriminating among the four classes of hadronic operators. This observation strongly motivates storage-ring experiments that can directly measure the EDMs of charged light nuclei. While EDMs of heavy diamagnetic atoms such as $^{199}$Hg, $^{129}$Xe, and $^{171}$Yb may provide suggestive information, they are not sufficient to draw definitive conclusions. Octupole-deformed systems such as $^{225}$Ra offer improved sensitivity compared with ordinary heavy atoms, but still provide less discriminatory power than light nuclei.

If the dominant hadronic UV CPV source is found to be either the QCD $\theta$ term or quark CEDMs, as indicated by the corresponding EDM ratios in Table~\ref{tab:ratios2}, this would also shed light on the origin of a nonzero QCD axion vacuum expectation value (VEV). Assuming that the strong CP problem is solved by the QCD axion, a $\bar{\theta}$-dominated scenario would suggest that the axion VEV arises primarily from PQ-breaking effects other than the QCD anomaly, such as quantum-gravitational effects. In contrast, in quark-CEDM-dominated scenarios, EDM ratios can first be used to determine the presence or absence of the Peccei--Quinn (PQ) mechanism. If the observed ratios are consistent with the predictions for quark CEDMs in the presence of PQ symmetry, this would indicate that the QCD axion VEV is predominantly induced by quark CEDMs in conjunction with the standard PQ breaking by the QCD anomaly.

The remaining two classes of UV CPV operators involving electrons, namely the electron EDM and CP-odd electron--down-quark interactions, can be independently tested through EDM measurements of paramagnetic systems, in particular polar molecules. Electron EDMs can be clearly distinguished from CP-odd electron--nucleon interactions by observing nonvanishing EDMs in ThO and at least one additional polar molecule, provided the experimental uncertainties are below approximately 10\%. If the dominant CPV source involves a nonzero CP-odd electron--nucleon coupling, its UV origin may be either the electron--down-quark interaction or long-distance hadronic effects induced by hadronic UV CPV sources. These two possibilities can be discriminated by the presence of correspondingly large EDMs in light nuclei or diamagnetic atoms, characterized by either $d_{n,p} \sim 10^{-13}\, e\,{\rm cm}\times C_S$ or $\bar{g}_{0,1} \sim C_S$.

It should be emphasized that many of the currently available theoretical calculations used to connect CPV operators at the QCD scale to experimentally measured EDMs are subject to intrinsic uncertainties that cannot be reliably quantified with the present level of theoretical understanding of the relevant QCD, nuclear, and atomic physics effects. Our results should therefore be interpreted with these uncertainties in mind. Our primary goal is to assess how far one can proceed using currently available theoretical inputs, while also motivating future theoretical efforts toward a more quantitative understanding of the matching relations relevant for EDMs.

In conclusion, future high-precision EDM measurements have the potential to provide profound insights into the underlying UV physics, allowing, in principle, the identification of SM contributions, key classes of BSM scenarios, the PQ mechanism, and the origin of PQ symmetry breaking. Further efforts to reduce theoretical uncertainties, particularly through lattice QCD calculations and complementary theoretical tools, are therefore strongly motivated.

\section*{Acknowledgments}

This work was supported by the Institute for Basic Science (IBS) under project code IBS-R018-D1. We thank Nodoka Yamanaka and Tae-Sun Park for helpful discussions.

\appendix

\section{ Na\"ive Dimensional Analysis for hadronic CP-odd quantities}

In this appendix, we estimate the hadronic IR CPV parameters 
$X_i^{\rm IR} \in \{d_n, d_p, \bar{g}_0, \bar{g}_1, C_1, C_2\}$ 
induced by the non-leptonic UV CPV parameters 
$X_i^{\rm UV} \in \{\bar{\theta}, w, \tilde{d}_q, d_q\}$ 
($q=u,d,s$), using naive dimensional analysis (NDA), in order to compare with the estimates presented in Sec.~\ref{sec:IRsrc}.

Applying the NDA rules of Refs.~\cite{Manohar:1983md,Georgi:1986kr,Weinberg:1989dx}, and taking into account chiral and isospin symmetries, we obtain
\bea
d_{n,p} &\sim& e\,\frac{m_*}{\Lambda_\chi^2}\,\bar{\theta}
+ e\,\tilde{d}_{u,d,s} + d_{u,d,s} + e f_\pi w, \label{NDA_dN}\\
\bar{g}_0 &\sim& \frac{m_*}{f_\pi}\,\bar{\theta}
+ 4\pi \Lambda_\chi\,\tilde{d}_{u,d}
+ (m_u+m_d)\Lambda_\chi\, w, \label{NDA_g0}\\
\bar{g}_1 &\sim& \frac{(m_u-m_d)}{m_s}\frac{m_*}{f_\pi}\,\bar{\theta}
+ 4\pi \Lambda_\chi (\tilde{d}_u-\tilde{d}_d)
+ (m_u-m_d)\Lambda_\chi\, w, \label{NDA_g1}\\
C_{1,2} &\sim& \frac{m_*}{f_\pi^2 \Lambda_\chi^2}\,\bar{\theta}
+ \frac{1}{f_\pi^2}(\tilde{d}_u+\tilde{d}_d)
+ \frac{1}{f_\pi}\, w, \label{NDA_C12}
\eea
evaluated at the matching scale $\mu_{\rm NDA}=225~\mathrm{MeV}$, defined by $\alpha_s(\mu_{\rm NDA})/(4\pi)\simeq 1/6$, where the one-loop QCD beta function becomes comparable to the two-loop contribution \cite{Weinberg:1989dx}. Here $m_* \equiv \left(\sum_{q=u,d,s} m_q^{-1}\right)^{-1} \simeq m_u m_d/(m_u+m_d)$ and $\Lambda_\chi = 4\pi f_\pi$. These expressions should be regarded as order-of-magnitude estimates with undetermined overall signs.

Several remarks are in order. First, for the definition of the quark CEDMs $\tilde{d}_q$, we adopt the convention of Eq.~(\ref{L2}), in which the QCD coupling $g_s(\mu_{\rm NDA})\sim 4\pi$ is absorbed into the operator. Second, the quark CEDM contribution to $\bar{g}_0$ is not necessarily proportional to $\mathrm{Tr}(\tilde{d}_q)=\tilde{d}_u+\tilde{d}_d$, since there exists an additional isospin-invariant contribution proportional to $\mathrm{Tr}(M_q^{-1}\tilde{d}_q)=\sum_q \tilde{d}_q/m_q$, arising from the effective CP-odd quark mass induced by $\tilde{d}_q$ \cite{Pospelov:2001ys}. NDA suggests that these two contributions can be comparable. However, Ref.~\cite{Pospelov:2001ys} shows, using QCD sum rules, that the contribution proportional to $\mathrm{Tr}(M_q^{-1}\tilde{d}_q)$ is exactly canceled by the shift of the QCD axion VEV in the presence of the Peccei--Quinn (PQ) mechanism. 

On the other hand, NDA indicates that, for $\bar{g}_1$, the contribution proportional to $\mathrm{Tr}(M_q^{-1}\tilde{d}_q)$ is negligible compared to that proportional to $\mathrm{Tr}(\tilde{d}_q \tau_3)$. Consequently, $\bar{g}_1$ is approximately proportional to $\tilde{d}_u-\tilde{d}_d$. Finally, in estimating $\bar{g}_1$ from $\bar{\theta}$, we use the fact from chiral perturbation theory that the leading contribution arises from $\eta$--$\pi^0$ mixing combined with the $\eta \bar{N}N$ coupling.

The expressions in Eqs.~(\ref{NDA_dN})--(\ref{NDA_C12}) cannot be directly compared with the results in Sec.~\ref{sec:IRsrc}, since the matching scales differ. While NDA uses $\mu_{\rm NDA}=225~\mathrm{MeV}$, most estimates in Sec.~\ref{sec:IRsrc} are given at $\mu=1~\mathrm{GeV}$. To translate between these scales, we use the renormalization group (RG) relations
\bea
d_q(1~\mathrm{GeV})/d_q(\mu_{\rm NDA}) &\sim& 1, \\
\tilde{d}_q(1~\mathrm{GeV})/\tilde{d}_q(\mu_{\rm NDA}) &\sim& 1, \\
m_q(1~\mathrm{GeV})/m_q(\mu_{\rm NDA}) &\sim& \frac{1}{2}, \\
g_s(1~\mathrm{GeV})/g_s(\mu_{\rm NDA}) &\sim& \frac{1}{2}, \\
w(1~\mathrm{GeV})/w(\mu_{\rm NDA}) &\sim& 2,
\eea
as obtained from the RG equations in Sec.~\ref{RGE}.

Using these relations, we find the NDA estimates at $\mu=1~\mathrm{GeV}$:
\bea
d_{n,p} &\sim& 3\times 10^{-3}\, e\,\mathrm{GeV}^{-1}\,\bar{\theta}
+ e\,\tilde{d}_{u,d,s} + d_{u,d,s}
+ 0.04\, e\,\mathrm{GeV}\, w, \\
\bar{g}_0 &\sim& 3.8\times 10^{-2}\,\bar{\theta}
+ 13\,\mathrm{GeV}\,\tilde{d}_{u,d}
+ 7\times 10^{-3}\,\mathrm{GeV}^2\, w, \\
\bar{g}_1 &\sim& 1.0\times 10^{-3}\,\bar{\theta}
+ 13\,\mathrm{GeV}(\tilde{d}_u-\tilde{d}_d)
+ 2.5\times 10^{-3}\,\mathrm{GeV}^2\, w, \\
C_{1,2} &\sim& 0.3\,\mathrm{GeV}^{-3}\,\bar{\theta}
+ 120\,\mathrm{GeV}^{-2}(\tilde{d}_u+\tilde{d}_d)
+ 10\,\mathrm{GeV}^{-1}\, w,
\eea
where $\tilde{d}_q$, $d_q$, and $w$ are evaluated at $\mu=1~\mathrm{GeV}$.

Overall, the NDA estimates are in agreement with the dedicated computations discussed in Sec.~\ref{sec:IRsrc} up to factors of $O(1)$. The main exceptions are $\bar{g}_0(\tilde{d}_q)$ and $\bar{g}_0^{\rm PQ}(\tilde{d}_q)$ in Eqs.~(\ref{gbar0_dqt}) and (\ref{gbar0_dqt'}), which are smaller than the NDA estimates by factors of $\sim 10$ and $\sim 5$, respectively. The factor $\sim 10$ suppression in $\bar{g}_0(\tilde{d}_q)$ can be partially attributed to cancellations between the contributions proportional to $\mathrm{Tr}(\tilde{d}_q)$ and $\delta m_N\,\mathrm{Tr}(M_q^{-1}\tilde{d}_q)$.

\bibliography{edm_ptc}
\bibliographystyle{utphys}

\end{document}